\documentclass{article}

\usepackage[preprint, nonatbib]{nips_2018}

\usepackage{float}

\usepackage[square,numbers]{natbib}
\usepackage[utf8]{inputenc} 
\usepackage[T1]{fontenc}    
\usepackage{hyperref}       
\usepackage{url}            
\usepackage{booktabs}       
\usepackage{amsfonts}       
\usepackage{nicefrac}       
\usepackage{microtype}      
\usepackage{graphicx}
\usepackage{tabularx}
\usepackage{hyperref}
\usepackage{amssymb}
\usepackage{amsmath}
\usepackage{array}
\usepackage{colortbl}
\usepackage{xcolor}
\usepackage{diagbox}
\usepackage{makecell}
\usepackage{subfigure}
\usepackage{longtable}

\title{Applying Back Propagation Algorithm and Analytic
	Hierarchy Process to Environment Assessment}

\author{
  Chunyu Sui
  \textsuperscript{1} \\
  \texttt{suichunyu@hotmail.com} \\
\And
  Xinrui Li
  \textsuperscript{1} \\
  \texttt{lixinrui@outlook.com} \\
\And
  Yinghang Song
  \textsuperscript{1} \\
  \texttt{songyinghang12138@outlook.com} \\
\And
  Chen Wu
  \textsuperscript{1} \\
  \texttt{wu\_chen@mail.sdu.edu.cn} \\
\And
  Ziyang Zhang
  \textsuperscript{3} \\
  \texttt{ziyangzhang@outlook.com} \\
\And
  School of Computer Science and Technology, Shandong University$^{1}$
\And
  School of Energy and Power Engineering, Shandong University$^{1}$
\And
  School of Computer Science and Technology, Shandong University$^{3}$
}

\begin{document}
	\maketitle

	\begin{abstract}
		This paper designs a new and scientific environmental quality 
		assessment method, and takes Saihan dam as an example to explore the 
		environmental
		improvement degree to the local and Beijing areas. AHP method is used 
		to assign
		values to each weight $7$ primary indicators and $21$ secondary 
		indicators were used
		to establish an environmental quality assessment model. The conclusion 
		shows
		that after the establishment of Saihan dam, the local environmental 
		quality has
		been improved by 7 times, and the environmental quality in Beijing has 
		been
		improved by $13\%$. Then , the future environmental index is predicted. 
		Finally the
		Spearson correlation coefficient is analyzed, and it is proved that 
		this method’s
		correlation is $99\%$. Finally, the back-propagation algorithm is used 
		to test and
		prove that the error is little.
	\end{abstract}

	\section{Introduction}
		\subsection{Problem Background}
			With the development of human society and human's claim on nature, ecological damage and environmental pollution have posed a great threat to the survival and development of human beings. Protecting and improving the ecological environment to achieve sustainable development of human society is the most urgent task for all mankind.
			
			The protection of ecological environment is to contribute to the sustainable development of human society. In recent years, people have built a green barrier against sand and dust storms, and there have been numerous activities to improve the ecological environment, such as afforestation, sand and water consolidation, etc.  People have gone on to devote themselves to improving ecological environment, advanced wave upon wave, which has also achieved great success. According to collected statistics, by the end of 2018, China's forest coverage had reached 22.96\%, with a cumulative forest area of 2.2 hectares, and the ratio keeps going up, making China the fastest-growing forest region in the world.
			\begin{figure}[H]
				\centering
				\includegraphics[width=\textwidth, height=6cm]{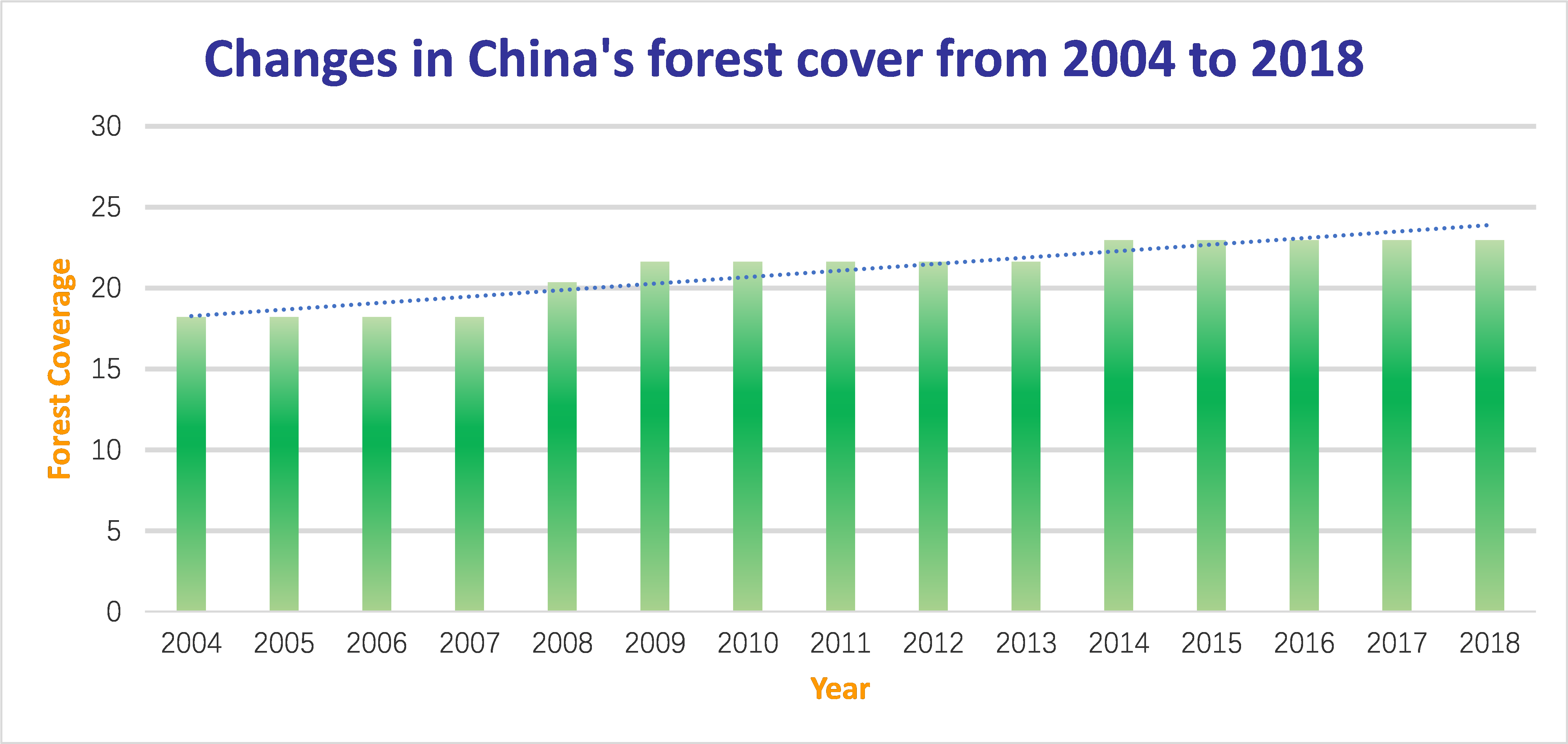}
				\caption{Changes in China's forest coverage from 2004 - 2018}
			\end{figure}
			
			The improvement in ecological environment will bring quantities of benefits to all creatures on Earth. By analyzing the weather data of cities around the reforested areas, we can get the first-hand information of changes in the urban ecological environment before and after the reforestation. At the same time, modeling different regions of China and the world, we can predict suitable locations for the establishment of ecological reservation, which is a great guide for the improvement of the global ecological environment.
		\subsection{Our work}
			Firstly, we collected the local ecological conditions now and before the establishment of the Saihanba Forest and then gave formulas to quantitatively assess the effect of the establishment of the Saihanba Forest on the local ecological environment.
			
			Secondly, we established the sandstorm risk index model and computed the coefficients of each weight by AHP algorithm. Then we quantified the sandstorm risk index of Beijing before and after the establishment of the Saihanba. Finally, we performed correlation tests using the Spearson Model.
			
			Thirdly, we selected Gansu Province as our research object and calculated the area of nature reserves to be established in Gansu province. For carbon emissions, we built a carbon storage model based on the carbon storage within the forest and calculated the annual average carbon storage of the newly established nature reserves.
			
			Then, we selected Mongolia, where the pollution level is more severe, for the study. We also quantified the pollution level of Gobi Sumbeer province and determined that the province needs to establish a nature reserve of 94,430 km2. Finally, we used a carbon storage model to assess its impact on greenhouse gases and carbon emissions, and finally concluded that the average annual carbon storage is 535,927,760 tons.
			
			Finally, we combined the established sandstorm risk index model and carbon storage model to suggest the establishment of the ecological model of Saihanba and the coordination of the ecosystem, respectively. Factors such as adjusting the structure of forest leaves and controlling environmental impacts are applied to the protection of ecological environment, so that sustainability can develop along with human civilization.
	\section{Preparation of the Models}
		\subsection{Assumptions and Justifications}
			$\bullet$ Nine indicators such as visibility and wind speed can be used to quantify the quality of the ecological environment.
			
			$\bullet$ Data from IKH nature reserves can be used for calculating data related to newly established nature reserves.
			
			$\bullet$ The collected data is real and reliable.
			
			$\bullet$ External conditions do not interfere with carbon neutrality.
		\subsection{Notations}
			The primary notations used in our paper are listed in Table 1.
			\begin{table}[H]
				\centering
				\begin{tabular}{p{6cm}<{\centering}p{6cm}<{\centering}}
					\Xhline{2pt}
					Symbol & Defination \\
					\Xhline{1pt}
					\rowcolor{blue!20}
					$EI$ & Ecosystem Status Index\\
					$CR$ & Consistency Ratio \\
					\rowcolor{blue!20}
					$CI$ & Consistency Indicator \\
					$RI$ & Stochastic Consistenct Index \\
					\rowcolor{blue!20}
					$H$ & Risk Degree of Sandstorm \\
					$DV$ & Visibility Indicator \\
					\rowcolor{blue!20}
					$U$ & Wind Index \\
					$\Delta T$ & Cooling Index \\
					\rowcolor{blue!20}
					$P$ & Sand Transport Index\\
					$TR$ & Sandstorm Development Trned Indicator\\
					\rowcolor{blue!20}
					$C$ & Carbon Storage of Different Forests \\
					$EH$ & Ecological Hazard Index \\
					\Xhline{2pt}
				\end{tabular}
			\end{table}
	\section{Proposed Model:}
		\subsection{Model}
			We collected information from the Saihanba Mechanical Forestry 
			Station and obtained data on relevant indicators. These indicators 
			are forested land area, forest cover, forest stock, live wood 
			stock, average annual number of windy days, average annual 
			precipitation, and frost-free period\cite{bib1,bib8}.
			
			Then, based on the Technical Specification of Ecological Environment Status Evaluation of the Ministry of Ecology and Environment of China, we define the composite indicator EI as the ecological environment score which can be calculated by the formula in the following:
			\begin{equation}
				EI = 62.42FC + 15.12FR + 4.51S + \frac{5.24D}{365} + \frac{3.04DF}{365} + 3.02RF
			\end{equation}
			
			$FC(hm^2)$: Area of mechanical forestry.
			
			$FR$: Forest cover ratio.
			
			$S$: Total Accumulation.
			
			$D$: Average annual number of high wind days.
			
			$DF$: Frost-free period.
			
			$EF$: Average annual precipitation.
		\subsection{Solutions}
			Firstly, in terms of area of mechanical forestry(FC), we queried and obtained the forest area before and after the establishment of Saihanba Forestry which are shown as follows:
			\begin{equation}
				Pre_s = 24
			\end{equation}
			\begin{equation}
				After_s = 115.1
			\end{equation}
			Then, according to the Technical Specification of Ecological Environment Status Evaluation of the State Ministry of Ecology and Environment, we speciated the data:
			\begin{equation}
				Pre = AIO \times 0.35 \times 24 / 140 = 30.675
			\end{equation}
			\begin{equation}
				After = AIO \times 0.35 \times 115 / 140 = 147.116
			\end{equation}
			where $AIO$ is 511.264.So. And we show the results in Fig \ref{fig6}:
			\begin{figure}[H]
				\centering
				\includegraphics[width=\textwidth, height=9cm]{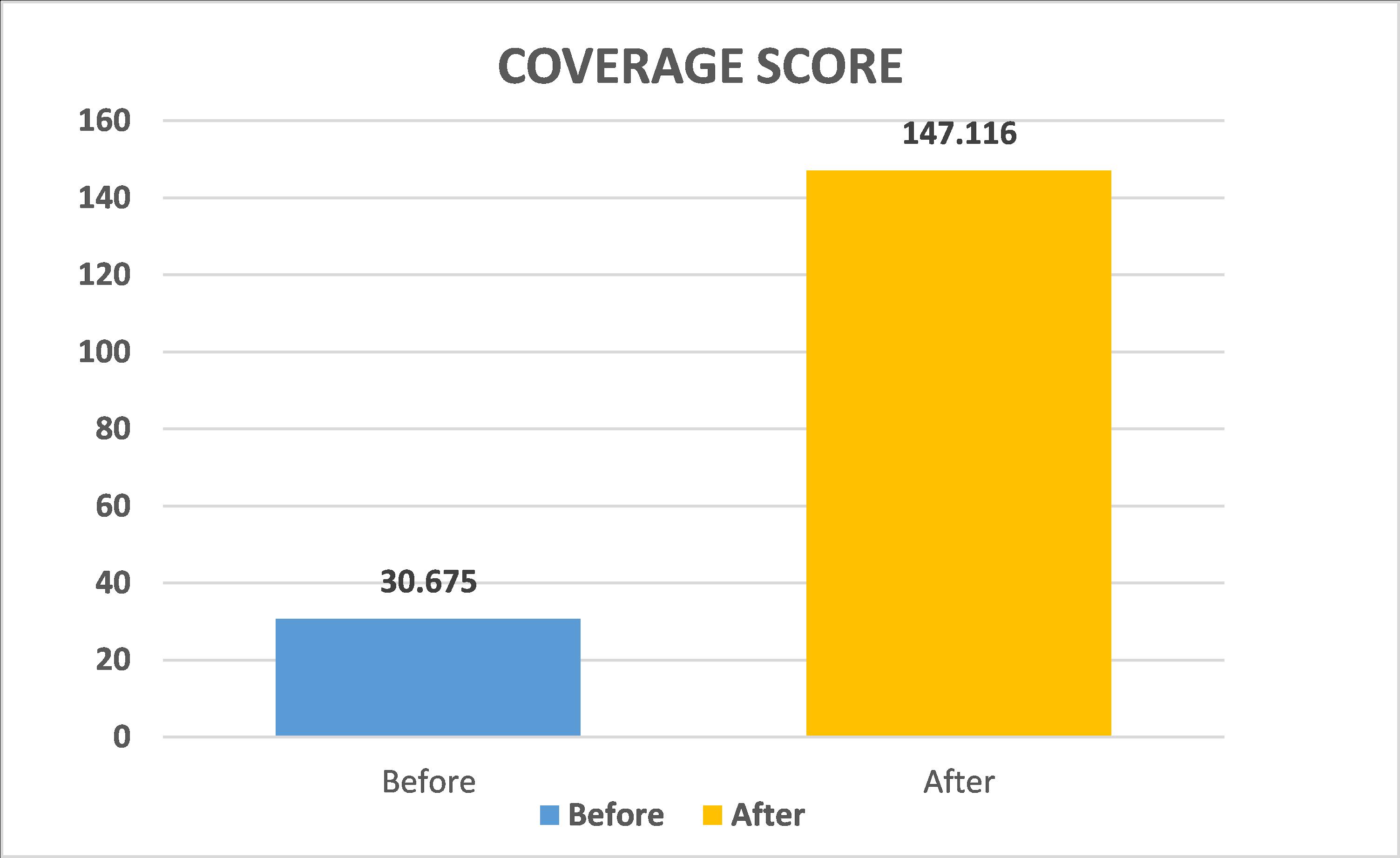}
				\caption{Forest Coverage}
				\label{fig6}
			\end{figure}
			According to the Technical Specification for the Evaluation of the 
			Ecological Environment Status of the Ministry of Ecology and 
			Environment of the State\cite{bib3}, we obtained the weighting 
			coefficients. Then, after calculation\cite{bib9}, we got the 
			score($F_{cs}$):
			\begin{equation}
				f_{cs} = \frac{A_1 \times F_c}{sq}
			\end{equation}
			the results are showing in Fig \ref{fig7}:
			\begin{figure}[H]
				\centering
				
				\includegraphics[width=\textwidth, height=9cm]{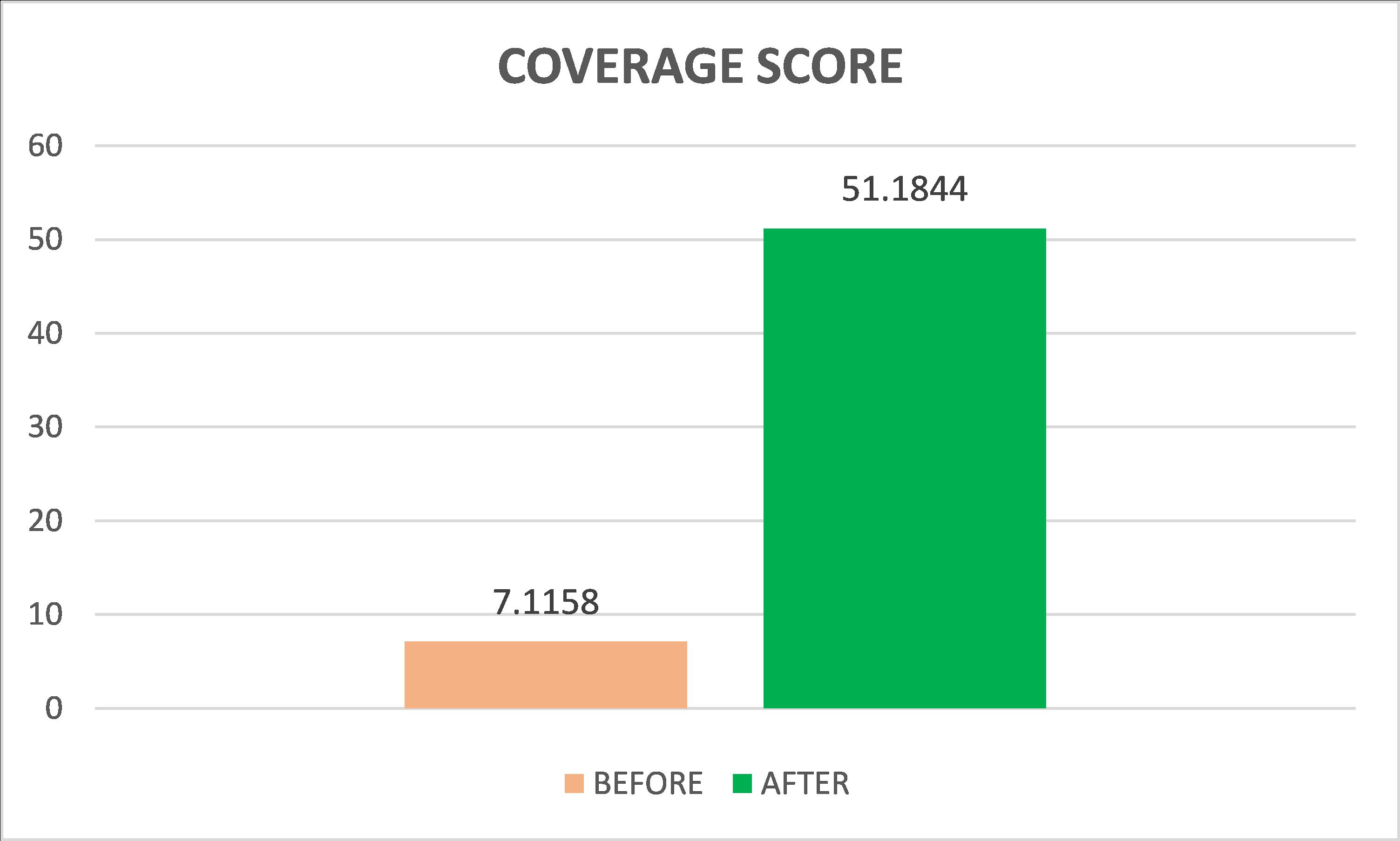}
				\caption{Forest Coverage Score}
				\label{fig7}
			\end{figure}
			It is calculated that the Saihanba forestry site has made a great 
			ecological achievement\cite{bib10,bib11}, and in the official 
			indicator given for the area of forest land In the past, it rose 
			sharply from 7.1158 to 51.1844 now, which is almost 7 times more.
			
			Using the same method, we can obtain the comparison of forest cover scores, which is shown is the following:
			\begin{figure}[H]
				\centering
				\includegraphics[width=\textwidth, height=9cm]{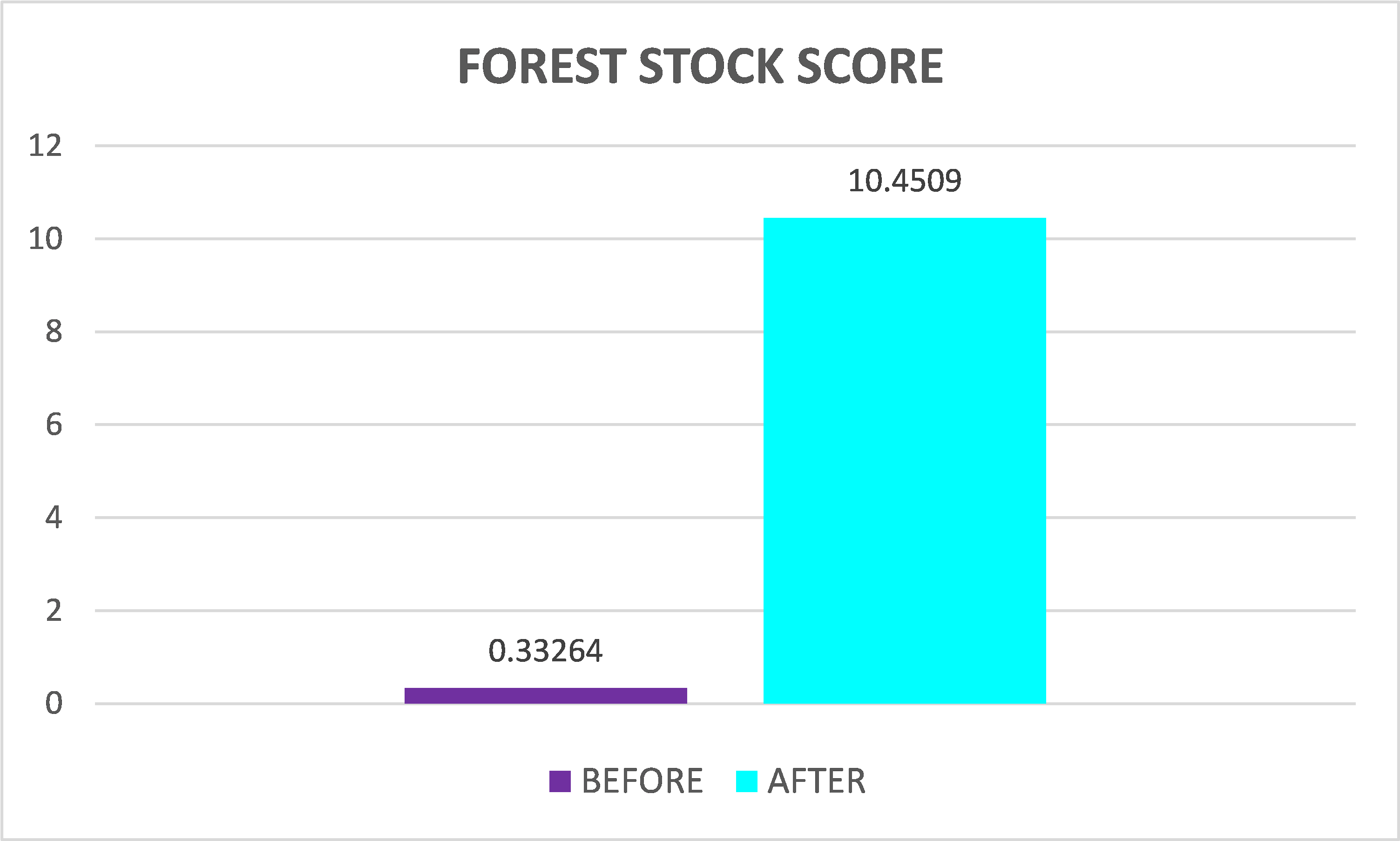}
				\caption{Forest Stock Score}
			\end{figure}
			From 0.33264 in the past to 10.4509 now, it has increased by $10.4509 / 0.33264 = 30.41$ times.
			
			Continuous, using the method above, We calculated all the metrics and represented them in Table:
			\begin{table}[H]
				\centering
				\caption{all the metrics}
				\begin{tabular}{p{1.5cm}<{\centering}p{1.5cm}<{\centering}p{1.5cm}<{\centering}p{1.5cm}<{\centering}p{1.5cm}<{\centering}p{1.5cm}<{\centering}p{1.5cm}<{\centering}}
					\Xhline{2pt}
					\diagbox{}{} & FC & FR & S & D & EF & DF \\
					\Xhline{1pt}
					\rowcolor{blue!20}
					Before & 0.17 & 0.114 & 0.096 & 0.46 & 0.891 & 7.21 \\
					After & 0.82 & 0.82 & 0.784 & 0.15 & 1.04 & 8.87 \\
					\Xhline{2pt}
				\end{tabular}
			\end{table}
			The final total calculated score is as follows:
			\begin{figure}[H]
				\centering
				\includegraphics[width=\textwidth, height=9cm]{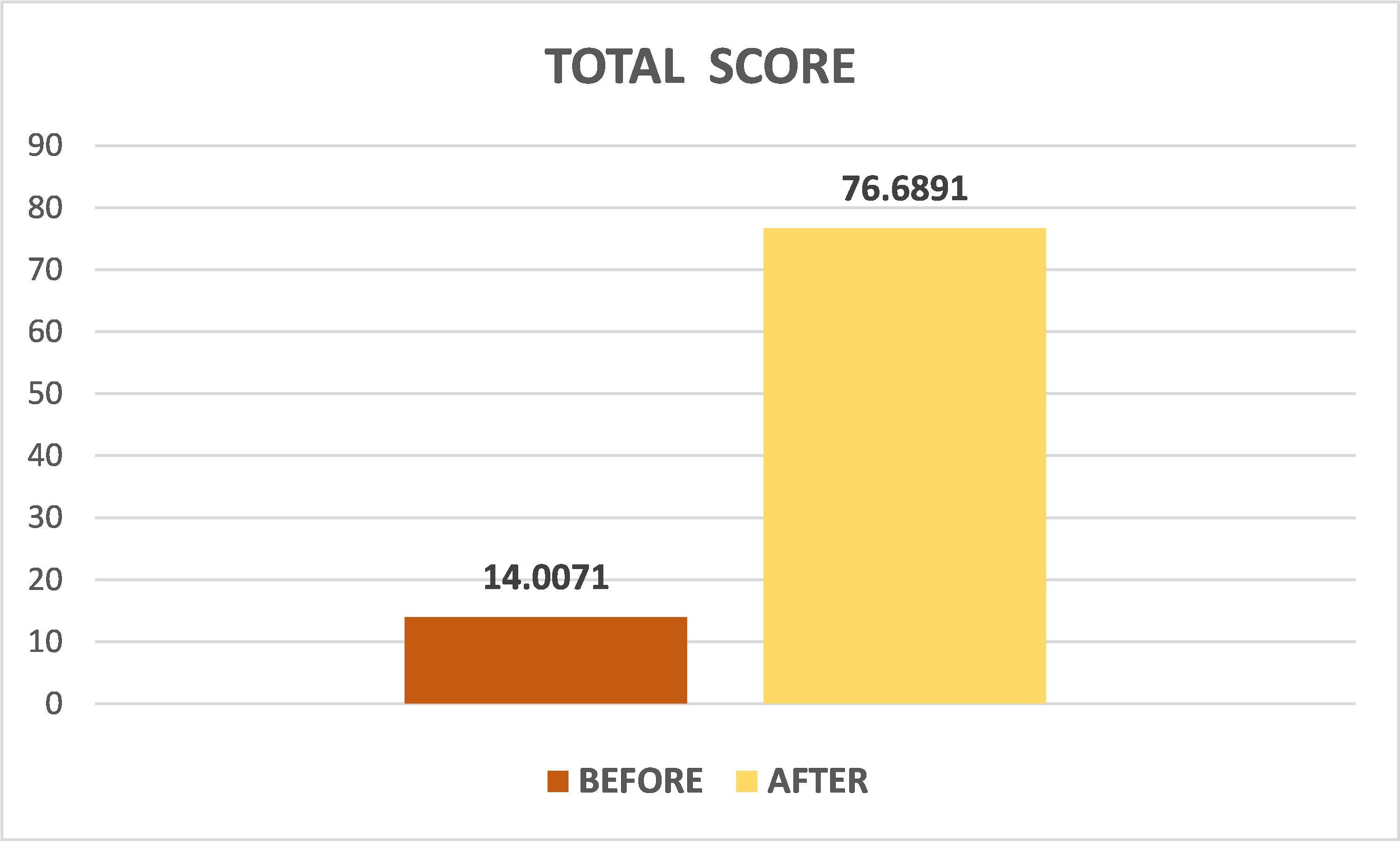}
				\caption{Total Score}
			\end{figure}
			The ecological score of Saihanba Forestry Reserve has made a great 
			leap from 14.0071 points in 1962 to 76.6891 points now. According 
			to the measure of protected area score of developed countries 
			released by National Bureau of Statistics\cite{bib12}, protected 
			areas with more than 48 points are considered outstanding nature 
			reserves\cite{bib13}. Saihanba has gone from a low score at the 
			early stage of its establishment to meeting the measure of 
			protected areas of developed countries, which is an outstanding 
			achievement of China in the field of nature conservation.
	\section{Weights Obtained}
		\subsection{Establishment of the Impact Evaluation System}
			\subsubsection{Analysis method selection}
	
				\noindent{$\divideontimes$ \bfseries Brief Introduction of AHP Method}
				
				There are many methods of data analysis, what we commonly use are methods such as AHP method, fuzzy method, TOPSIS method, etc\cite{bib2}. In this paper, we use AHP method to analyse Saihanba's impact on Beijing's ability on sandstorm resistance. The general ideas of AHP method are as follows:
				
				\hspace{1sp} (1) Analyze the hierarchy and structure of the target problem. Then clarify and construct a network of the relationships between the levels.
				
				\hspace{1sp} (2) Use the scaling method to construct the judgement matrix. The matrix is constructed layer by layer from the lower level to the higher level. In this paper, we choose Saaty 1-9 scale method, and details are showing in table \ref{tab1}:
				\begin{table}[H]
					\centering
					\caption{The Saaty 1-9 scale method}
					\label{tab1}
					\begin{tabular}{cp{10cm}<{\centering}}
						\Xhline{2pt}
						Scales & Events \\
						\Xhline{1pt}
						\rowcolor{blue!15}
						1 & Indicates that the two factors are equally important. \\
						3 & One factor is slightly more important than another. \\
						\rowcolor{blue!15}
						5 & One factor is significantly more important another. \\
						7 & One factor is strongly more important than another. \\
						\rowcolor{blue!15}
						9 & One factor is extremely more important  than another factor. \\
						2, 4, 6, 8 & The median value in the above adjacency judgment \\
						\rowcolor{blue!15}
						Countdown & \begin{tabular}[c]{@{}c@{}}If the value of i compared with j is $a_{ij}$,\\ then the value of j compared with i is 1/a. \end{tabular} \\
						Negative numbers & \begin{tabular}[c]{@{}c@{}}If the Pearson correlation coefficient between i and j is negative,\\ then $a_{ij}$ is negative. \end{tabular} \\
						\Xhline{2pt}
					\end{tabular}
				\end{table}
				\hspace{1sp} (3) Results consistency test. If the consistency ratio $CR$ < 0.1, then the consistency is considered to satisfy the allowable accuracy. The consistency ratio can be calculated by the following formula:
				\begin{equation}
					CR = CI / RI
				\end{equation}
				\begin{equation}
					CI = (\lambda - n) / (n - 1)
				\end{equation}
				In the formula, while CI represents consistency indicator, $\lambda$ represents the maximum eigenvalue of the judgement matrix, and n is dimension of the judgement matrix. If CI = 0, then there is absolute consistency in the results; If CI is close to 0, the results have satisfactory agreement.
				
				Also, RI denotes stochastic consistency index, and its values are showing in table \ref{tab2}:
				\begin{table}[H]
					\centering
					\caption{Values of RI}
					\label{tab2}
					\begin{tabular}{p{0.2cm}<{\centering}p{0.9cm}<{\centering}p{0.9cm}<{\centering}p{0.9cm}<{\centering}p{0.9cm}<{\centering}p{0.9cm}<{\centering}p{0.9cm}<{\centering}p{0.9cm}<{\centering}p{0.9cm}<{\centering}p{0.9cm}<{\centering}p{0.9cm}<{\centering}}
						\Xhline{2pt}
						n & 1 & 2 & 3 & 4 & 5 & 6 & 7 & 8 & 9 & 10 \\
						\Xhline{1pt}
						\rowcolor{blue!15}
						RI & 0 & 0 & 0.58 & 0.9 & 1.12 & 1.24 & 1.32 & 1.41 & 1.45 & 1.49 \\
						\Xhline{2pt}
					\end{tabular}
				\end{table}
				AHP method quantitative process is not objective enough, we add Pearson correlation coefficient in the analysis to help build a more scientific and reasonable judgment matrix.
				
				\noindent{$\divideontimes$ \bfseries Maximum Eigenvalue Approximation Method}
				
				When finding the maximum eigenvalue, to avoid finding all eigenvalues and then finding the maximum eigenvalue, we use the summation method to simplify the process. The steps are as follows:
				
				(1) Let $\boldsymbol{A} = (a_{ij})$ be a square matrix of order n. Then normalize all columns of $\boldsymbol{A}$ to obtain matrix $\boldsymbol{B} = (b_{ij})$ finally;
				
				(2) Sum matrix $\boldsymbol{B}$ row by row to obtain vector $\boldsymbol{C} = (C_1, C_2, \cdots, C_n)^T$;
				
				(3) Normalize $\boldsymbol{C}$ to obtain vector $\boldsymbol{W} = (W_1, W_2, \cdots, W_n)^T$;
				
				(4) According to the formula below to obtain the maximum eigenvalue $\lambda_{max}$:
				\begin{equation}
					\lambda_{max} = \frac{1}{n}\sum\limits_{i = 1}^{n}\frac{(\boldsymbol{A}\boldsymbol{W})_i}{W_i}
				\end{equation}
				where n is the approximate eigenvector.
				
				\noindent{$\divideontimes$ \bfseries Data Positivization and Dimension Removal}
				
				When performing data processing, positive dimensionless data is required, so we use the threshold method to nondimensionalize the original data, the formula is as follows:
				\begin{equation}
					\label{equ1}
					a_{ij new} = \frac{a_{j max} + a_{j min} - a_{ij}}{a_{j max}}
				\end{equation}
				
				In the formula \ref{equ1}, $a_{j max}$ is the maximum value of all the values in column j. Likewise, $a_{j min}$ is the minimum value of all the values in column j. Meanwhile, $a_{ij}$ represents the current data, and $a_{ij new}$ represents the new data after nondimensionalizing.
			\subsubsection{Index System}
				We collected weather data of Beijing from 1956 to 2021, and selected March - May, when dust storms are frequent in each year, as the research object, and used temperature, visibility, wind speed, sea level pressure, precipitation, dew point difference, and 24-hour temperature difference\cite{bib2} as the measurement indexes to establish the evaluation system shown in Fig \ref{fig1}:
				\begin{figure}[H]
					\centering
					\includegraphics[width=\textwidth, height=12cm]{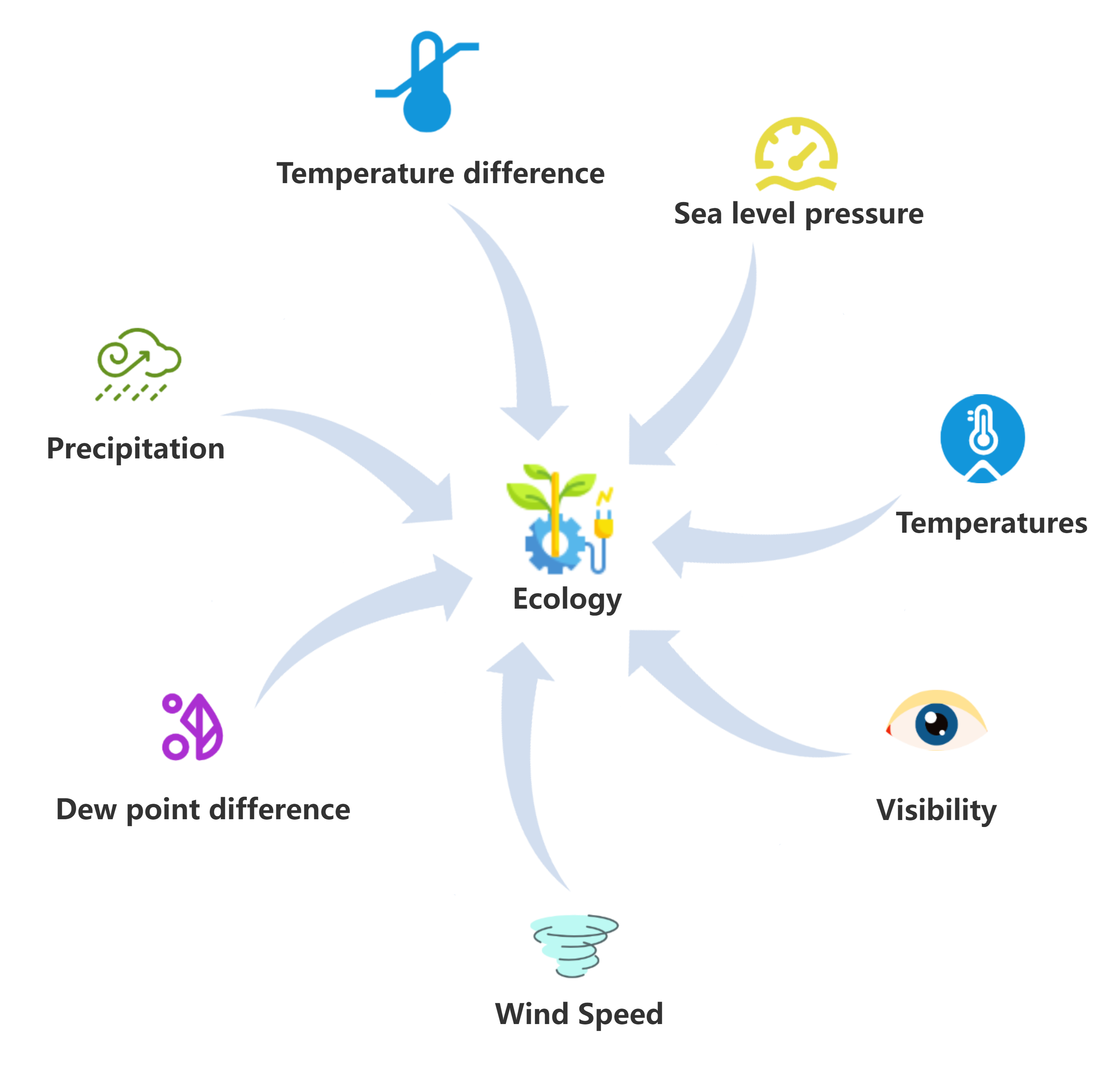}
					\caption{The Evaluation System}
					\label{fig1}
				\end{figure}
				Considering the magnitude of the influence of different indicators on the resistance to dust storms and the simplicity of the formula we choose five influencing factors: Visibility, Wind Index, Cooling Index, Sand Transport Index and Trend Indicator. Meanwhile, to enhance the holistic nature of metrics, we gave the weight occupied by each indicator separately according to the magnitude of the influence of each indicator. This resulted in the establishment of a comprehensive index that can reflect the influence of the Saihanba mechanical forest on the wind and sand resistance of Beijing called Risk Degree of sandstorms(abbreviated as H).
				
				Then, we make the hierarchical disgram to demonstrate the structure of AHP method:
				\begin{figure}[H]
					\centering
					\includegraphics[width=\textwidth, height=10cm]{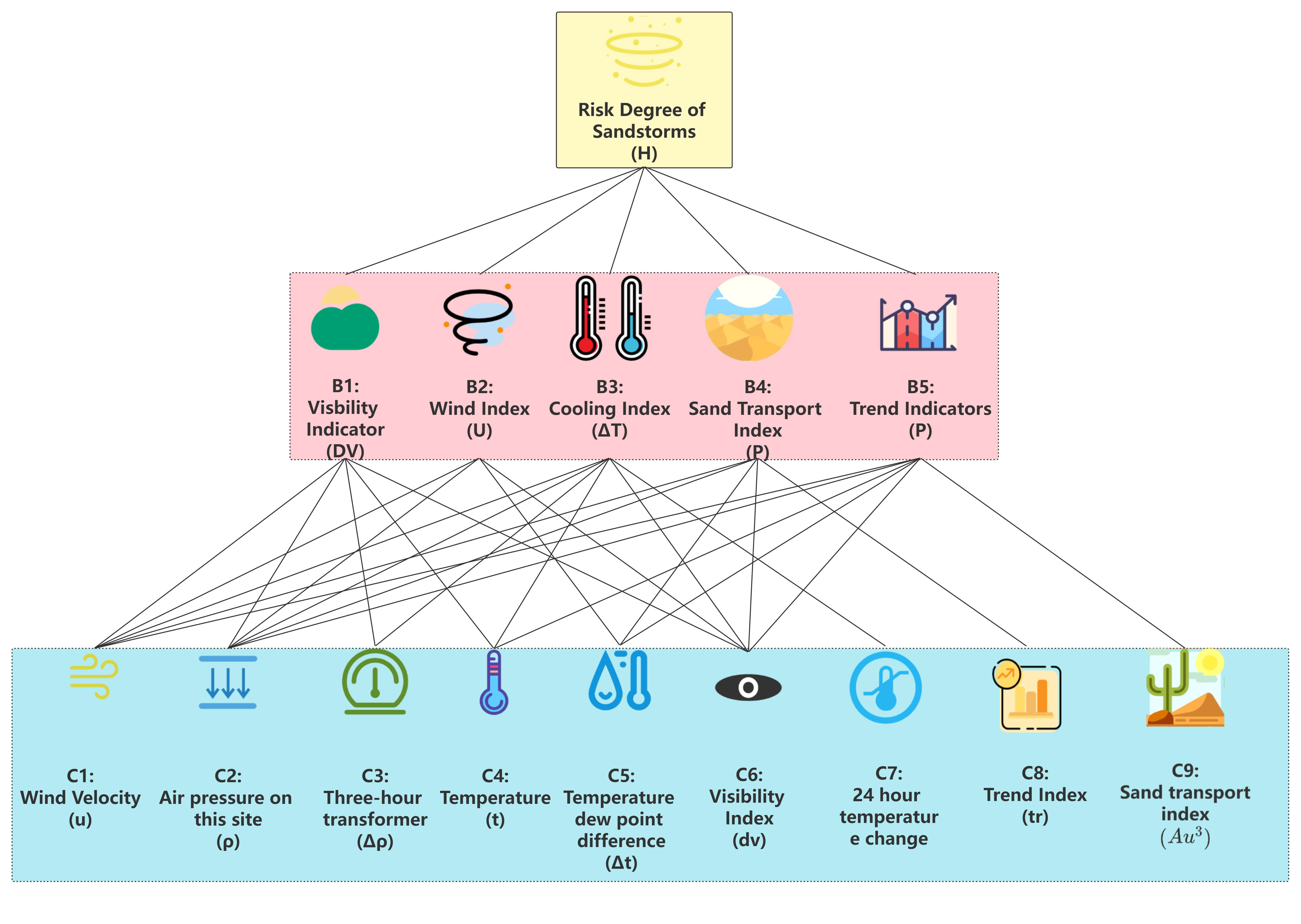}
					\caption{The structure of the evaluation system}
					\label{fig2}
				\end{figure}
				As we can see from Fig \ref{fig2}, we use the factors in layer C to build the judgment matrix for the indicators in layer B, and finally get the composite score based on the relationship between the indicators and the indicators' weights.
		\subsection{Calculation of the Indicators}
			Risk Degree of Sandstorms Expression includes 5 known indicators and 1 unknown indicator. The following is a description of how each indicator is calculated.
			
			\vspace{2mm}
			\noindent{$\divideontimes$ \bfseries Calculation of Visibility Index(DV)}
			
			\begin{equation}
				DV = dv + x_1
			\end{equation}
			where $dv$ is visibility index, and $x_1$ can be concluded from the correlation analysis between meteorological factors.
			
			Visibility Index is determined by visibility and human eye recognition target contrast thresholds. It can be computed by the following equations:
			\begin{equation}
				V = \frac{1}{a}\ln \frac{1}{\varepsilon}
			\end{equation}
			\begin{equation}
				\begin{aligned}
					dv &= \ln a \approx \frac{3.912}{a}
				\end{aligned}
			\end{equation}
			\begin{equation}
				dv = \ln a
			\end{equation}
			where $V$ is visibility and $\varepsilon$ is human eye recognition target contrast thresholds.
			
			\vspace{2mm}
			\noindent{$\divideontimes$ \bfseries Calculation of Wind Index(U)}
			
			In this context, the wind which we discussed here means wind at ground level. In that case, the Wind Index(U) can be calculated by the formula as follows:
			\begin{equation}
				U = u + x_1
			\end{equation}
			where $u$ is the wind velocity at ground level, and $x_2$ represents meteorological factors related to wind speed.
			
			\vspace{2mm}
			\noindent{$\divideontimes$ \bfseries Calculation of Cooling Index($\Delta$T)}
			
			Cooling Index reflects the change of temperature during the day.
			\begin{equation}
				\Delta T = \Delta t_{24} - (\Delta t_{24})_{min} + x_3
			\end{equation}
			where $\Delta t_{24}$ is denotes the temperature change in the last 24 hours, and $(\Delta t_{24})_{min}$ is the minimum temperature change in the last 24 hours. $x_3$ is uncertainty factor.
			
			\vspace{2mm}
			\noindent{$\divideontimes$ \bfseries Calculation of Sand Transport Index(P)}
			
			\begin{equation}
				P = A u^3 + x_4
			\end{equation}
			where $A$ represents lower bedding surface factor; $u^3$ is the cube of wind velocity also called sand transport factor. Meanwhile, $x_4$ denotes the uncertainty of other factors.
			
			\vspace{2mm}
			\noindent{$\divideontimes$ \bfseries Calculation of Sandstorm Development Trend Indicator(TR)}
			
			\begin{equation}
				\begin{aligned}
					TR &= tr + x_4 \\
					&= u - u_s + x_4
				\end{aligned}
			\end{equation}
			where $tr$ is sand storm trend values; $u$ and $u_s$ represents wind speed and wind speed of sand initiation respectively.
		\subsection{Calculation Process}
			\noindent {$\divideontimes$ \bfseries Judgement Matrix}
			
			After reading the article and scoring the indicators, we obtain the judgement matrix of each  indicator. The results are as follows:
			
			The judgement matrix of Visibility Indicator(DV):
			\begin{equation}
				\begin{array}{cc}
					B1 & \begin{array}{ccccc} C6 & \hspace{4mm}C1 & \hspace{4mm}C2 & \hspace{4mm}C3 & \hspace{4mm}C4 	\end{array} \\
					\begin{array}{c} C6 \\ C1 \\ C2 \\ C3 \\ C4 \end{array} & 
					\begin{bmatrix}
						0.236 & 0.279 & 0.234 & 0.187 & 0.223 \\
						0.170 & 0.201 & 0.214 & 0.270 & 0.236 \\
						0.192 & 0.179 & 0.190 & 0.240 & 0.210 \\
						0.196 & 0.175 & 0.186 & 0.155 & 0.136 \\
						0.206 & 0.166 & 0.177 & 0.147 & 0.195 \\
					\end{bmatrix}
				\end{array}
			\end{equation}
			thus, we obtained the maximum eigenvalue of matrix B1: $\lambda = 5.137$, $CI = 0.034$ and $CR = 0.031$. Since CR < 0.1, the judgment matrix passes the consistency test. And the weight vector of B1 $\vec{n_1}$ is 
			\begin{equation}
				\vec{n_1} = (0.218, 0.202, 0.170, 0.178, 0.232, 0, 0, 0)^T
			\end{equation}
			
			The judgement matrix of Wind Index(U):
			\begin{equation}
				\begin{array}{cc}
					B2 & \begin{array}{cccc} C1 & \hspace{4mm}C2 & \hspace{4mm}C5 & \hspace{4mm}C6 \end{array} \\
					\begin{array}{c} C1 \\ C2 \\ C5 \\ C6 \end{array} & 
					\begin{bmatrix}
						0.296 & 0.336 & 0.279 & 0.275 \\
						0.214 & 0.242 & 0.263 & 0.265 \\
						0.244 & 0.212 & 0.230 & 0.231 \\
						0.246 & 0.210 & 0.228 & 0.229 \\
					\end{bmatrix}
				\end{array}
			\end{equation}
			thus, we obtained the maximum eigenvalue of matrix B2: $\lambda = 4.0097$, $CI = 0.0032$ and $CR = 0.00359$. Since CR < 0.1, the judgment matrix passes the consistency test. And the weight vector of B2 $\vec{n_2}$ is 
			\begin{equation}
				\vec{n_2} = (0.296, 0.246, 0, 0, 0.229, 0.228, 0, 0, 0)^T
			\end{equation}
			
			The judgement matrix of Cooling Index($\Delta$ T):
			\begin{equation}
				\begin{array}{cc}
					B3 & \begin{array}{cccccc} C7 & \hspace{4mm}C1 & \hspace{4mm}C2 & \hspace{4mm}C3 & \hspace{4mm}C4 & 	\hspace{4mm}C6 \end{array} \\
					\begin{array}{c} C7 \\ C1 \\ C2 \\ C3 \\ C4 \\ C6 \end{array} & 
					\begin{bmatrix}
						0.211 & 0.216 & 0.178 & 0.173 & 0.158 & 0.365 \\
						0.152 & 0.156 & 0.163 & 0.165 & 0.167 & 0.126 \\
						0.172 & 0.138 & 0.145 & 0.146 & 0.149 & 0.112 \\
						0.175 & 0.136 & 0.142 & 0.143 & 0.146 & 0.110 \\
						0.185 & 0.129 & 0.135 & 0.136 & 0.138 & 0.104 \\
						0.105 & 0.225 & 0.236 & 0.238 & 0.242 & 0.182 \\
					\end{bmatrix}
				\end{array}
			\end{equation}
			thus, we obtained the maximum eigenvalue of matrix B3: $\lambda = 6.1172$, $CI = 0.0234$ and $CR = 0.0189$. Since CR < 0.1, the judgment matrix passes the consistency test. And the weight vector of B3 $\vec{n_3}$ is 
			\begin{equation}
				\vec{n_3} = (0.155, 0.144, 0.142, 0.138, 0, 0.205, 0.217, 0, 0)^T
			\end{equation}
			
			The judgement matrix of Sand Transport Index(P):
			\begin{equation}
				\begin{array}{cc}
					B4 & 
					\begin{array}{ccccc} 
						C8 & \hspace{4mm}C1 & \hspace{4mm}C2 & \hspace{4mm}C5 & \hspace{4mm}C6 		\end{array} \\
					\begin{array}{c} 
						C8 \\ C1 \\ C2 \\ C5 \\ C6 
					\end{array} & 
					\begin{bmatrix}
						0.258 & 0.408 & 0.202 & 0.199 & 0.249 \\
						0.129 & 0.204 & 0.275 & 0.276 & 0.259 \\
						0.213 & 0.124 & 0.167 & 0.167 & 0.157 \\
						0.215 & 0.122 & 0.165 & 0.166 & 0.156 \\
						0.186 & 0.141 & 0.191 & 0.191 & 0.180 \\
					\end{bmatrix}
				\end{array}
			\end{equation}
			thus, we obtained the maximum eigenvalue of matrix B4: $\lambda = 5.112$, $CI = 0.028$ and $CR = 0.025$. Since CR < 0.1, the judgment matrix passes the consistency test. And the weight vector of B4 $\vec{n_4}$ is 
			\begin{equation}
				\vec{n_4} = (0.229, 0.165, 0, 0, 0.165, 0.178, 0, 0.263, 0)^T
			\end{equation}
			
			The judgement matrix of Sand Transport Index(P):
			\begin{equation}
				\begin{array}{cc}
					B5 & \begin{array}{cccccc} C9 & \hspace{4mm}C1 & \hspace{4mm}C2 & \hspace{4mm}C4 & \hspace{4mm}C5 & \hspace{4mm}C6 \end{array} \\
					\begin{array}{c} C9 \\ C1 \\ C2 \\ C4 \\ C5 \\ C6 \end{array} & 
					\begin{bmatrix}
						0.211 & 0.356 & 0.178 & 0.161 & 0.169 & 0.225 \\
						0.108 & 0.181 & 0.232 & 0.236 & 0.234 & 0.219 \\
						0.172 & 0.113 & 0.145 & 0.148 & 0.147 & 0.137 \\
						0.183 & 0.107 & 0.137 & 0.140 & 0.138 & 0.129 \\
						0.178 & 0.110 & 0.141 & 0.143 & 0.142 & 0.133 \\
						0.149 & 0.132 & 0.168 & 0.171 & 0.170 & 0.158 \\
					\end{bmatrix}
				\end{array}
			\end{equation}
			thus, we obtained the maximum eigenvalue of matrix B5: $\lambda = 6.109$, $CI = 0.0218$ and $CR = 0.0176$. Since CR < 0.1, the judgment matrix passes the consistency test. And the weight vector of B5 $\vec{n_5}$ is 
			\begin{equation}
				\vec{n_5} = (0.202, 0.144, 0, 0.139, 0.141, 0.158, 0, 0, 0.217)^T
			\end{equation}
			
			The importance of each sub-indicator in the composite index is different, so we refered to the literature\cite{bib1} and gave the weights of the sub-indicators. Then we created Table\ref{tab3} to show the weights:
			\begin{table}[H]
				\centering
				\caption{Weights of each indicators}
				\label{tab3}
				\begin{tabular}{p{1.7cm}<{\centering}p{1.7cm}<{\centering}p{2cm}<{\centering}p{2cm}<{\centering}p{2cm}<{\centering}p{2cm}<{\centering}}
					\Xhline{2pt}
					Indicators & Visibility Index($DV$) & Wind Index($U$) & Cooling Index($\Delta T$) & Sand Transport Index($P$) & Development Trend Indicator($TR$) \\
					\Xhline{1pt}
					\rowcolor{blue!15}
					Weights & $0.159$ & $0.403$ & $0.088$ & $0.077$ & $0.274$ \\
					\Xhline{2pt} 
				\end{tabular}
			\end{table}
			According to Table \ref{tab3}, the greatest influence on the danger of sandstorms in Beijing is wind speed, accounting for 0.403 of the total degree of influence, followed by the trend indicator TR, indicating that the trend of dust storms also has a greater influence on dust storms, visibility indicators have a dramatic influence on the danger, while the factors with less influence are sand transport indicators and cooling indicators.
			\begin{figure}[H]
				\centering
				\includegraphics[width=10cm, height=10cm]{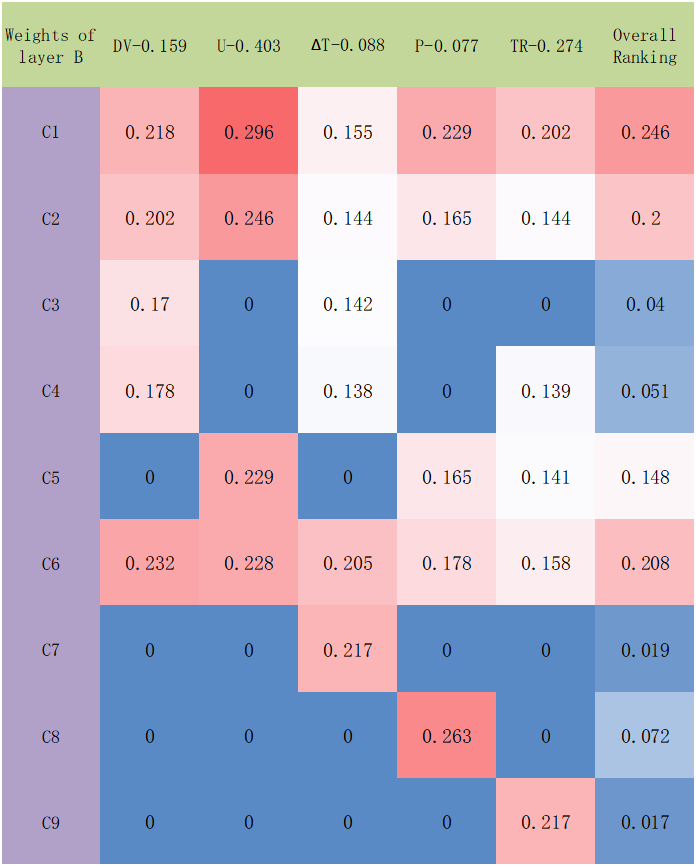}
				\caption{Overall Ranking of Sand Storm Danger in Beijing area}
				\label{fig8}
			\end{figure}
			From Fig \ref{fig8}, the formula for H can be obtained as:
			\begin{equation}
				\label{equ2}
				\begin{aligned}
					H &= f(DV, U, \Delta T, TR, P) \\
					&= 0.159 DV + 0.403 U + 0.088 \Delta T + 0.077 TR + 0.274 P
				\end{aligned}
			\end{equation}
			
			\vspace{2mm}
			\noindent {$\divideontimes$ \bfseries Data substitution}
			
			As can be seen from euqation \ref{equ2}, in Beijing, the most influential meteorological factor for dust storms is wind speed, with a weight of 0.246, followed by visibility, which are the classical factors for dust storm research, but there are other high-impact factors in the region, such as pressure at this site and temperature dew point difference, both of which have a weight of more than 0.14, and the degree of influence cannot be ignored.
			
			Taking the temperature indicator as an example, according to the above calculation method, we collected 20877 data in days for Beijing from 1956 to 2021, some of which are as follows:
			\begin{table}[H]
				\centering
				\begin{minipage}[H]{0.4\textwidth}
					\centering
					\caption{Part of the Data in 1956}
					\begin{tabular}{cc}
						\Xhline{2pt}
						DATE  & Celsius temperature \\
						\Xhline{1pt}
						\rowcolor{blue!20}
						1956-10-1 & 15.4398 \\
						1956-10-2 & 11.0556 \\
						\rowcolor{blue!20}
						1956-10-3 & 11.7678 \\
						1956-10-4 & 14.4359 \\
						\rowcolor{blue!20}
						1956-10-5 & 15.2574 \\
						1956-10-6 & 15.7259 \\
						\rowcolor{blue!20}
						1956-10-7 & 13.5874 \\
						1956-10-8 & 17.0215 \\
						\rowcolor{blue!20}
						1956-10-9 & 17.5784 \\
						1956-10-10 & 10.3539 \\
						\Xhline{2pt}
					\end{tabular}%
				\end{minipage}
				\begin{minipage}[H]{0.4\textwidth}
					\centering
					\caption{Part of the Data in 2020}
					\begin{tabular}{cc}
						\Xhline{2pt}
						\multicolumn{1}{c}{DATE} & \multicolumn{1}{c}{Celsius temperature} \\
						\Xhline{1pt}
						\rowcolor{blue!20}
						2020-3-1 & 2.9854 \\
						2020-3-2 & 3.5542 \\
						\rowcolor{blue!20}
						2020-3-3 & 1.6745 \\
						2020-3-4 & 1.1965 \\
						\rowcolor{blue!20}
						2020-3-5 & 2.8745 \\
						2020-3-6 & 5.0598 \\
						\rowcolor{blue!20}
						2020-3-7 & 8.8265 \\
						2020-3-8 & 4.7458 \\
						\rowcolor{blue!20}
						2020-3-9 & 5.3856 \\
						2020-3-10 & 4.2487 \\
						\Xhline{2pt}
					\end{tabular}
				\end{minipage}
			\end{table}
			For the temperature metric, we measure in annual units of measure. The average temperature\\($AVG_T$) values are obtained as follows:
			\begin{table}[H]
				\centering
				\caption{Average Temperature of 2012-2021}
				\begin{tabular}{p{1cm}<{\centering}p{1cm}<{\centering}p{0.8cm}<{\centering}p{0.8cm}<{\centering}p{0.8cm}<{\centering}p{0.8cm}<{\centering}p{0.8cm}<{\centering}p{0.8cm}<{\centering}p{0.8cm}<{\centering}p{0.8cm}<{\centering}p{1cm}<{\centering}}
					\Xhline{2pt}
					DATE  & 2012  & 2013  & 2014  & 2015  & 2016  & 2017  & 2018  & 2019  & 2020  & 2021 \\
					\Xhline{1pt}
					\rowcolor{blue!20}
					$AVG_T$ & 14.274 & 13.190 & 16.150 & 15.232 & 15.368 & 16.258 & 15.082 & 15.416 & 14.724 & 14.763 \\
					\Xhline{2pt}
				\end{tabular}
			\end{table}
			The detailed information is in Fig \ref{fig9}:
			\begin{figure}[H]
				\centering
				\includegraphics[width=\textwidth, height=8cm]{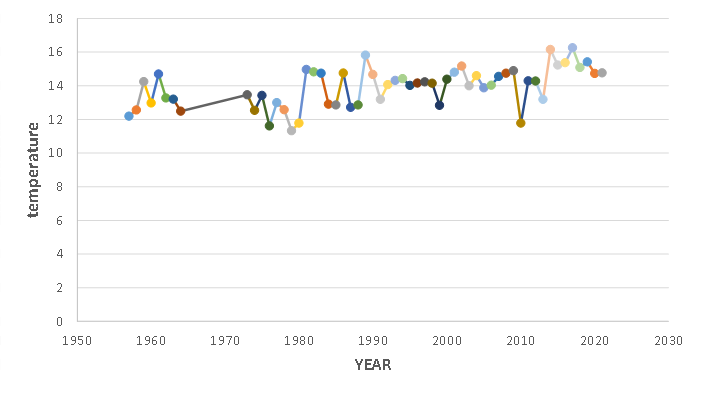}
				\caption{Average Temperature}
				\label{fig9}
			\end{figure}
			Then, for the visibility indicators that directly reflect the ability to resist dust storms, we also collected and counted the relevant results in the same way as Average Temperature. The results are as follows:
			\begin{table}[H]
				\centering
				\caption{Visibility}
				\begin{tabular}{p{1cm}<{\centering}p{1cm}<{\centering}p{0.8cm}<{\centering}p{0.8cm}<{\centering}p{0.8cm}<{\centering}p{0.8cm}<{\centering}p{0.8cm}<{\centering}p{0.8cm}<{\centering}p{0.8cm}<{\centering}p{0.8cm}<{\centering}p{1cm}<{\centering}}
					\Xhline{2pt}
					DATE  & 2012  & 2013  & 2014  & 2015  & 2016  & 2017  & 2018  & 2019  & 2020  & 2021 \\
					\Xhline{1pt}
					\rowcolor{blue!20}
					V & 4.968 & 6.719 & 6.550 & 6.385 & 6.284 & 6.305 & 5.259 & 5.715 & 5.971 & 5.562 \\
					\Xhline{2pt}
				\end{tabular}
			\end{table}
			The detailed information is in Fig \ref{fig10}:
			\begin{figure}[H]
				\centering
				\includegraphics[width=\textwidth, height=9cm]{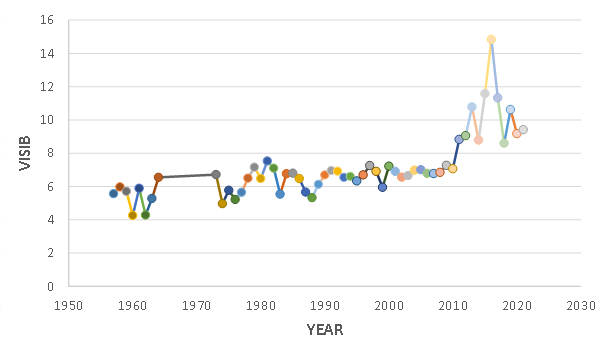}
				\caption{Visbility}
				\label{fig10}
			\end{figure}
			We used a time series analysis, and obtained the following visible capacity of Beijing for the next 10 years(PDC).
			\begin{table}[H]
				\centering
				\caption{Predictions}
				\begin{tabular}{p{1cm}<{\centering}p{1cm}<{\centering}p{0.8cm}<{\centering}p{0.8cm}<{\centering}p{0.8cm}<{\centering}p{0.8cm}<{\centering}p{0.8cm}<{\centering}p{0.8cm}<{\centering}p{0.8cm}<{\centering}p{0.8cm}<{\centering}p{1cm}<{\centering}}
					\Xhline{2pt}
					DATE  & 2022  & 2023  & 2024  & 2025  & 2026  & 2027  & 2028  & 2029  & 2030  & 2031 \\
					\Xhline{1pt}
					\rowcolor{blue!20}
					PDC & 11.518 & 10.194& 12.634 & 10.755 & 12.691 & 12.504 & 13.608 & 10.105 & 13.066 & 9.471\\
					\Xhline{2pt}
				\end{tabular}
			\end{table}
			\begin{figure}[H]
				\centering
				\includegraphics[width=\textwidth, height=8cm]{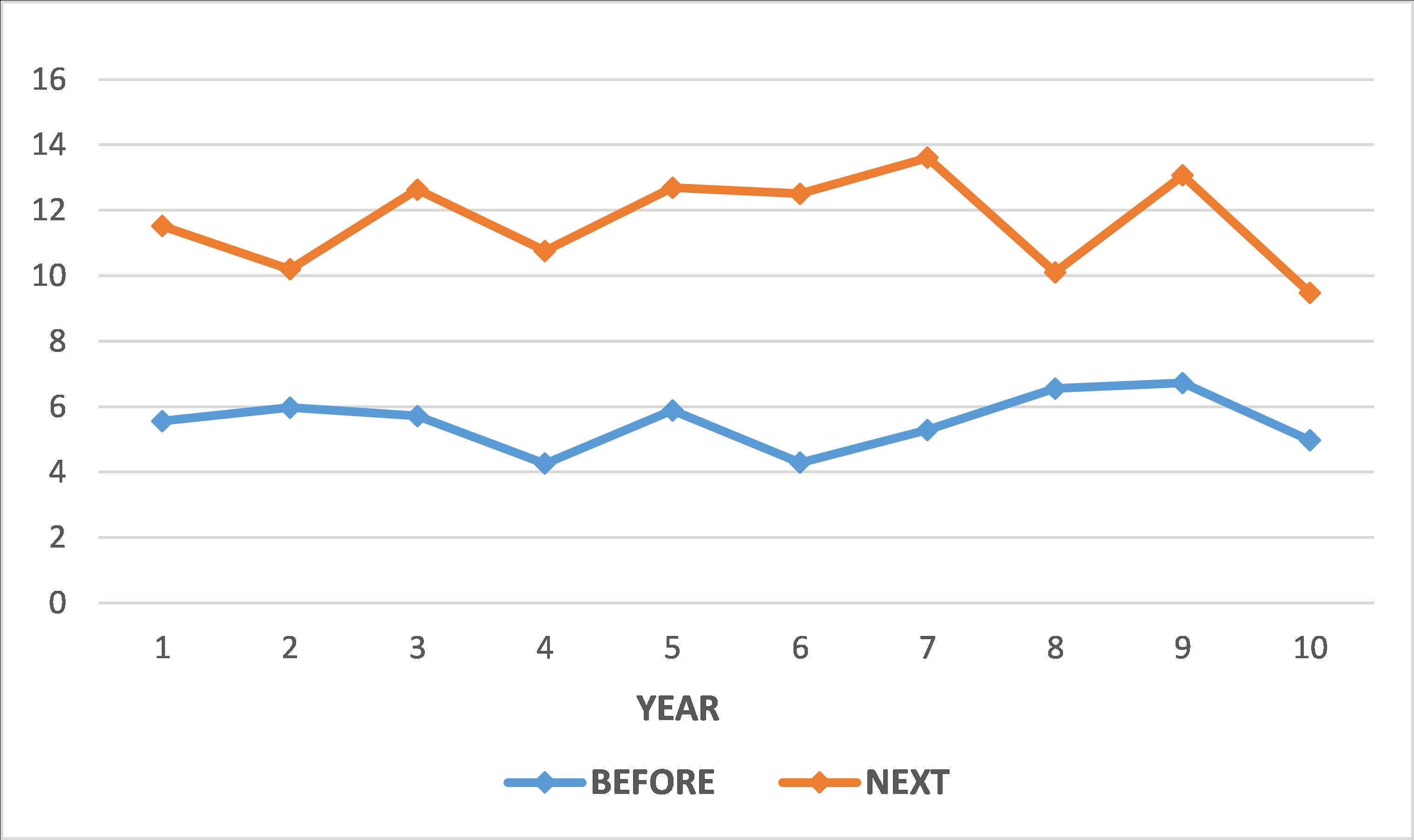}
				\caption{Visbility Comparison}
			\end{figure}
			Repeating the above steps, we can derive the annual statistics of Beijing on these five indicators.
			
			\vspace{2mm}
			\noindent {$\divideontimes$ \bfseries Conclusion}
			
			We can conclude that after the establishment of the Saihanba in 1962, the visibility of Beijing in March-May, when dust storms are frequent, has steadily increased and has been greatly improved, with 2010 as the dividing line. Meanwhile, the visibility capability will continue to grow at a high acceleration in the coming decade, which is inseparable from China's ecological efforts and strong investment in the Saihanba.
	\section{Calculation of Ecological Environment Indicators:}
		\subsection{Choose Geographical Locations}
			At the outset, we take account of the differences in China's geographical locations and economic conditions, so that each province and city has its own development level of ecological civilization construction. Then, combined with the descriptions in the Bulletin on the State of China's Ecological Environment, we classify domestical regions into three types of zones: good, moderate and poor, according to the ecological condition\cite{bib4}.
			
			Aftering classifying the regions, we selected the following regions: 
			
			(1) Sichuan and Yunnan, where ecological environment is good;
			
			(2) Liaoning, Guangxi and Shandong, where ecological environment is moderate;
			
			(3) Inner Mongolia, Gansu and Heibei. While Gansu located in the northwest, Heibei is more developed in heavy industry and lags behind in ecological protection. Their ecological environment is poor.
	
			\subsubsection{Solutions to the Model}
				So, firstly, we chose Gansu Province as our research object. Gansu Province located in the northwest of China which has complex and diverse landscapes and surrounded by mountains. As China has neglected that protecting ecological environment should go in parallel with economic construction, the ecological environment of Gansu Province has suffered a lot.
				
				According to the data collected from the air qualiity monitoring stations which constructed in Gansu Province, we got the average concentrations including $PM_{2.5}$ and $PM_{10}$ and, $CO$, $NO_2$, $SO_2$ and $O_3$ values in 11 study areas. The detail data is in the following:
				\begin{table}[H]
					\centering
					\caption{Average Concentration}
					\begin{tabular}{p{2.2cm}<{\centering}p{1.5cm}<{\centering}p{1.5cm}<{\centering}p{1.5cm}<{\centering}p{1.5cm}<{\centering}p{1.5cm}<{\centering}p{1.2cm}<{\centering}}
						\Xhline{2pt}
						City & $PM_{2.5}$ $(\mu g/m^3)$ & $PM_{10}$ $(\mu g/m^3)$ & $CO$ $(mg/m^3)$ & $NO_2$ $(\mu g/m^3)$ & $SO_2$ $(\mu g/m^3)$ & $O_3$ \\
						\Xhline{1pt}
						\rowcolor{blue!20}
						Jiayuguan & 26.53 & 78.26 & 0.68 & 20.61 & 18.25 & 70 \\
						Zhangye & 26.51 & 59.48 & 0.47 & 15.11 & 13.01 & 73 \\
						\rowcolor{blue!20}
						Jinchang & 21.40 & 72.50 & 0.93 & 15.14 & 26.34 & 73 \\
						Wuwei & 27.30 & 71.50 & 1.76 & 23.22 & 7.07 & 65 \\
						\rowcolor{blue!20}
						Baiyin & 25.32 & 70.76 & 0.74 & 25.73 & 35.06 & 44 \\
						Lanzhou & 38.80 & 98.80 & 0.94 & 42.59 & 11.34 & 58 \\
						\rowcolor{blue!20}
						Dinxi & 22.41 & 49.42 & 0.47 & 17.94 & 6.30 & 55 \\
						Pingliang & 36 & 71.42 & 0.74 & 40.80 & 8.60 & 65 \\
						\rowcolor{blue!20}
						Qingyang & 23.78 & 54.81 & 0.81 & 11.54 & 5.39 & 53 \\
						Tianshui & 16.41 & 40.37 & 0.57 & 18.15 & 9.40 & 63 \\
						\rowcolor{blue!20}
						Longnan & 24.90 & 61.13 & 0.60 & 18.42 & 13.06 & 65 \\
						\Xhline{2pt}
					\end{tabular}
				\end{table}
				In Lanzhou City, for example, the PM 2.5 level has been maintained at a high level.
				\begin{figure}[H]
					\centering
					\includegraphics[width=\textwidth, height=10cm]{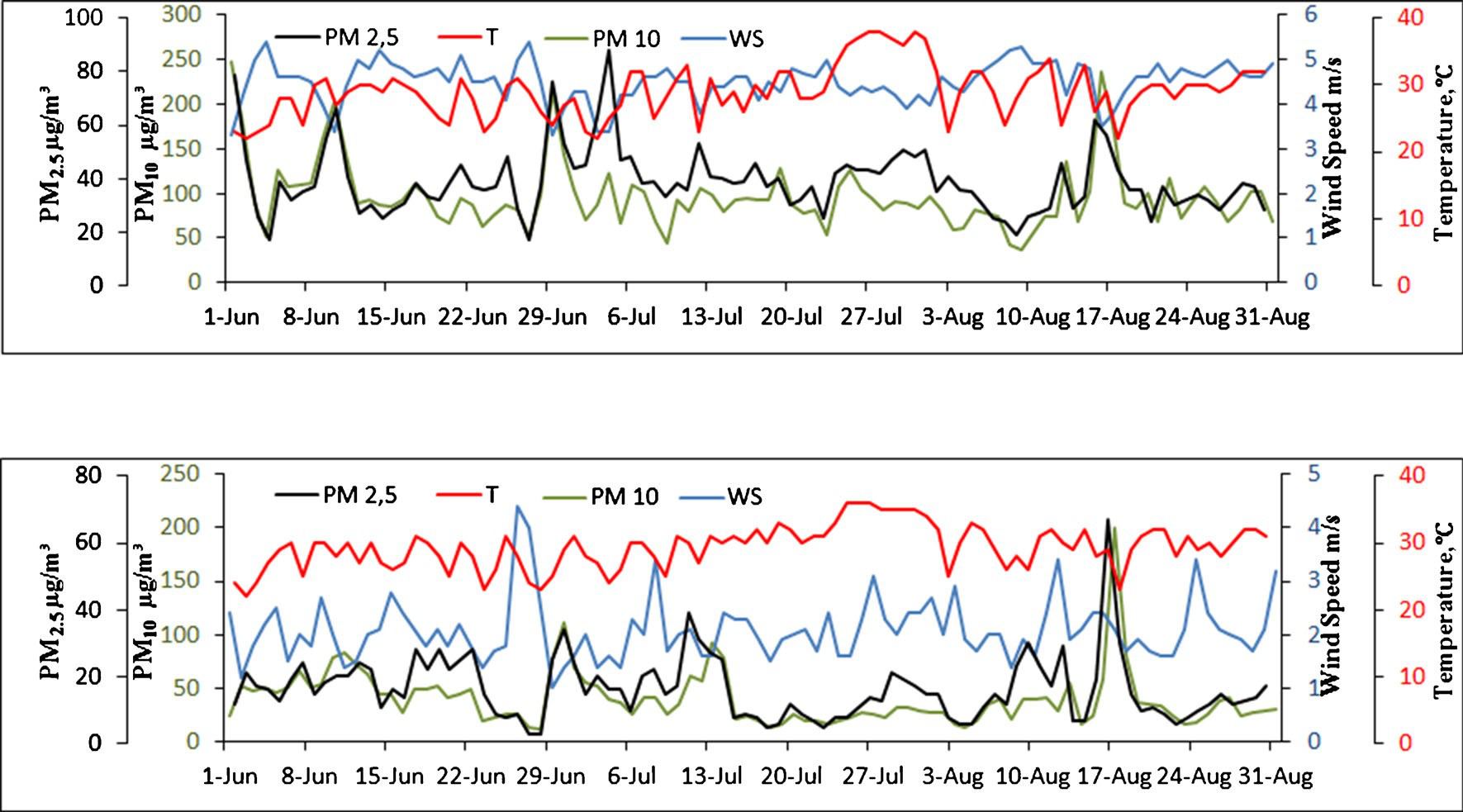}
					\caption{$PM_{2.5}$ level of Lanzhou City}
				\end{figure}
				\begin{figure}[H]
					\centering
					\includegraphics[width=\textwidth, height=5cm]{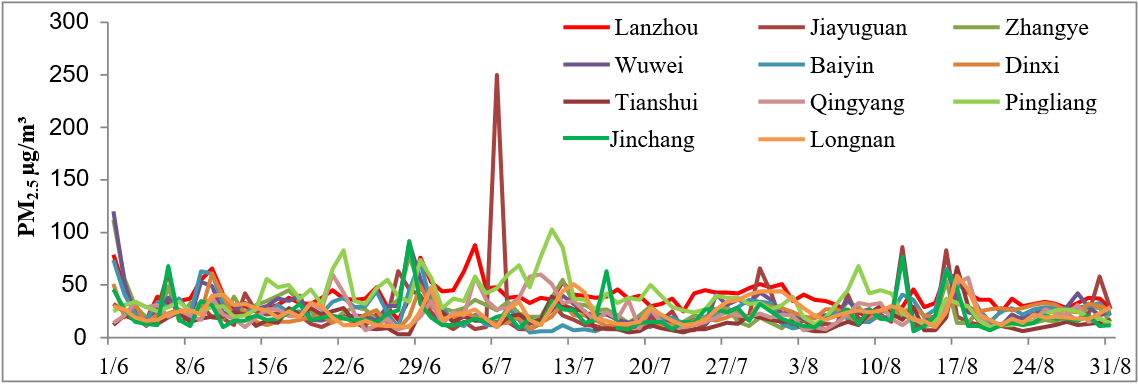}
					\caption{Comprehensive Average Concentrations of Lanzhou City}
				\end{figure}
				In summary, Gansu Province is a good target for our study. We collected data from Gansu Province for the past 5 years, and used 9 factors such as dew point (DEWP), wind speed (WSPD), rainfall (PRCR), weather station pressure (STP), sea level pressure (SLP), visibility (VISIB), 24-hour temperature difference (delta\_t), dust storm trend value (tr), and 3-hour dew point difference (delta\_dewp) as measurement factors to quantitatively analyze the ecological environmental protection in Gansu Province. Taking June 2020 as an example, some data are as follows.
				\begin{table}[H]
					\centering
					\caption{Quantitative Assessment Data}
					\begin{tabular}{p{1cm}<{\centering}p{1cm}<{\centering}p{1cm}<{\centering}p{1cm}<{\centering}p{1cm}<{\centering}p{1cm}<{\centering}p{1cm}<{\centering}p{1cm}<{\centering}p{1cm}<{\centering}p{1cm}<{\centering}}
						\Xhline{2pt}
						DATE  & DEWP  & VISIB & WDSP  & denta\_stp & denta\_slp & denta\_dewp & denta\_t & $u^3$  & tr \\
						\Xhline{1pt}
						\rowcolor{blue!20}
						2020-6-1 & 30    & 18.6  & 3.4   & 0.013 & 0.5   & 9.5   & 41.7221 & 117.649 & 0.9 \\
						2020-6-2 & 27    & 18.6  & 5.5   & 0.034 & 5.3   & 3.0   & 44.6110 & 405.224 & 3.4 \\
						\rowcolor{blue!20}
						2020-6-3 & 21.3  & 18.6  & 5.7   & 0.017 & 3.1   & 5.7   & 46.1110 & 1442.897 & 7.3 \\
						2020-6-4 & 31.6  & 18.6  & 4.5   & 0.004 & 0.3   & 10.3  & 45.6666 & 157.464 & 1.4 \\
						\rowcolor{blue!20}
						2020-6-5 & 37.6  & 18.6  & 5.0   & 0.031 & 3.9   & 6.0   & 47.2221 & 1092.727 & 6.3 \\
						2020-6-6 & 35.4  & 18.3  & 6.3   & 0.021 & 2.9   & 2.2   & 43.3332 & 857.375 & 5.5 \\
						\rowcolor{blue!20}
						2020-6-7 & 36.5  & 18.6  & 5.0   & 0.079 & 15.1  & 1.1   & 33.5554 & 1030.301 & 6.1 \\
						\Xhline{2pt}
					\end{tabular}
				\end{table}
				Using the above equation, we calculate the 24-hour temperature difference($\Delta$t), the dust storm trend value(tr), and the 3-hour dew point difference($\Delta$dewp). Some of the data are as follows:
				\begin{table}[H]
					\centering
					\caption{Values of $\Delta t, tr, \Delta dewp$(part of the data)}
					\begin{tabular}{p{2cm}<{\centering}p{2.5cm}<{\centering}p{2.5cm}<{\centering}p{2.5cm}<{\centering}p{2.5cm}<{\centering}}
						\Xhline{2pt}
						Year  & Month & $\Delta$ T    & $\Delta$ dewp & \multicolumn{1}{c}{tr} \\
						\Xhline{1pt}
						\rowcolor{blue!20}
						2020  & 1     & 15.0590 & 2.1903 & 1.7839 \\
						2020  & 2     & 19.8505 & 3.0172 & 4.3414 \\
						\rowcolor{blue!20}
						2020  & 3     & 25.4820 & 4.0677 & 4.4645 \\
						2020  & 4     & 33.4073 & 4.8200 & 6.3700 \\
						\rowcolor{blue!20}
						2020  & 5     & 38.2024 & 5.0935 & 5.5516 \\
						2020  & 6     & 42.6629 & 4.5833 & 4.2500 \\
						\rowcolor{blue!20}
						2020  & 7     & 44.9300 & 3.5839 & 3.7387 \\
						2020  & 8     & 42.6576 & 3.8613 & 3.7935 \\
						\rowcolor{blue!20}
						2020  & 9     & 37.3332 & 4.5200 & 2.5733 \\
						2020  & 10    & 28.2508 & 3.0097 & 3.6226 \\
						\rowcolor{blue!20}
						2020  & 11    & 20.5832 & 2.6733 & 2.6533 \\
						2020  & 12    & 11.7651 & 2.9677 & 1.7742 \\
						\Xhline{2pt}
					\end{tabular}
				\end{table}
				Then, we substitute into the formula \ref{equ2}, and the ratings obtained for the last 5 years are as follows:
				\begin{table}[H]
					\centering
					\caption{The ratings}
					\label{tab5}
					\begin{tabular}{p{2.8cm}<{\centering}ccccccc}
						\Xhline{2pt}
						Year & 2015 & 2016 & 2017 & 2018 & 2019 & 2020 & 2021 \\
						\Xhline{1pt}
						\rowcolor{blue!20}
						Ecological Warning Index & 42.7296 & 39.5110 & 39.6897 & 41.4805 & 42.5098 & 40.3942 & 39.8290 \\
						\Xhline{2pt}
					\end{tabular}
				\end{table}
				It can be seen from \ref{tab5} that the ecological warning index of Gansu Province is much higher than 20 which is the warning index given in the "Provincial Ecological Civilization Construction Evaluation Report", so Gansu Province needs and urgently needs to establish nature reserves.
			\subsubsection{Sizing, assessing carbon emissions}
				We take Yanchi Bay National Nature Reserve in Gansu Province as an example to quantitatively illustrate the impact of establishing nature reserves on environmental protection efforts.
				
				Yanchi Bay National Nature Reserve was established in 2006 and is part of Jiuquan. Among the data we have collected on the ecological status of Subei County from 2001 to 2020, part of the data for 2021 are as follows:
				
				Among them, we obtained the Ecological Index for the last 20 years according to the adventure formula in TASK 2, and we show it in Fig \ref{fig3} as follows:
				\begin{figure}[H]
					\centering
					\includegraphics[width=\textwidth, height=8cm]{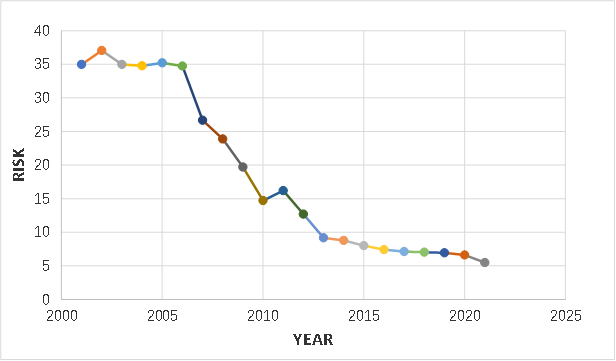}
					\caption{The Ecological Index for the Last 20 Years}
					\label{fig3}
				\end{figure}
				It can be seen that Jiuquan's ecological condition index has 
				risen nearly five times after the establishment of the Yanchi 
				Bay National Nature Reserve\cite{bib19}. So, combining TASK 2 
				and the data of this question\cite{bib20}, we can conclude that 
				the establishment and size of nature reserves are positively 
				correlated with the decrease of ecological risk index.
				
				The area of Jiuquan is 192,000 square kilometers, and the area of Yanchi Bay National Nature Reserve is 63,000 square kilometers. Menwhile, the area of Gansu Province is 453,700 square kilometers.
	
				Then, we measure the ecological conservation capacity of nature reserves by the upward value of the ecological condition index per unit area($\theta$). Taking Yanchi Bay Nature Reserve as an example, the upward value of ecological risk index per unit area($\theta$) is
				\begin{equation}
					\theta = 19 \times 192000 / 63000 = 48.761
				\end{equation}
				Then, the ecological condition index that can rise in 10 years for 1 square meter in the nature reserve is 48.761. Likewise, Gansu Province covers an area of 453,700 square kilometers, and the average value of the Ecological Index in the territory reached 40.213. If you want the ecological condition index of Gansu Province to be no less than 20 percent of the warning line in 10 years' time, then the area of nature reserve to be established($S_q$) is:
				\begin{equation}
					S_q = (40.213 - 20) \times 453700 / 10 = 188073.21629
				\end{equation}
				
				Therefore, it was concluded that 4 nature reserves of $50,000 km^2$ would need to be established in order not to fall below the ESI alert line after 10 years.
		\subsection{Evaluate Its Impact on Absorbing Greenhouse Gases and Mitigating Carbon Emissions}
			\subsubsection{Calculation of Carbon Stocks in Different Forests}
				In a large forest, there may be hundreds or thousands of tree species. Here we give the calculation method of carbon stock for different kinds of woods.
				
				\vspace{2mm}
				\noindent{$\divideontimes$ \bfseries Carbon Stock of Arbor Forest }
				
				Firstly, based on forest stock, we use the forest stock expansion method to calculate tree biomass through the stock expansion coefficient. Then, using the data collected, we obtained the dry weight of biomass from the volumetric density (or dry weight coefficient). Finally, the carbon sequestration is calculated from the carbon content rate. The formulas along the calculation process are in the following:
				\begin{equation}
					C_{arbor} = \gamma \sum\limits_{i = 1}^{n}V_i(WD)_i(BEF)_i(1 + R)
				\end{equation}
				
				$C_{arbor}(/t \times 10^4)$: Carbon sequestration by forest trees.
				
				$\gamma$: Carbon conversion factor, international common value is 0.5.
				
				$V_i$: Stumpage volume, weight of wood volume of all types of standing trees existing in a given area of forest.
				
				$WD_I(/t \cdot m^{-1})$: Wood base density.
				
				$BEF_i$: Biological expansion factor, dimensionless value, take IPPC default value 1.3.
				
				$R$: Rootstock Ratio, dimensionless value, take IPPC default value 0.42.
				The variables and their meanings are showing in the table below:
	
				\vspace{2mm}
				\noindent{$\divideontimes$ \bfseries Carbon Stock of Economic Forest}
				
				Economic forest carbon stock is the product of economic forest biomass and carbon content. The calculation formula is:
				\begin{equation}
					C_{economic} = W_{economic} \cdot A_{economic} \cdot (CF)_{economic}
				\end{equation}
				
				$C_{economic}$: Economic forest carbon stocks.
				
				$W_{economic}(t \cdot hm^{-2})$: Biomass per unit area of economic forest, average value is $23.70\ t \cdot hm^{-2}$.
				
				$A_{economic}(hm^2)$: Area of economic forests.
				
				$(CF)_{economic}$: Carbon content of economic forests, taken as 0.47.
				
				\vspace{2mm}
				\noindent{$\divideontimes$ \bfseries Carbon Stock of Bamboo Forest}
				
				Bamboo forest carbon stock is the product of bamboo forest biomass and carbon content. The calculation formula is:
				\begin{equation}
					C_{bamboo} = W_{bamboo} \cdot N_{bamboo} \cdot (CF)_{bamboo}
				\end{equation}
	
				$C_{bamboo}$: Carbon stock of bamboo forest.
				
				$W_{bamboo}(kg/(per\ plant))$: Average biomass per plant in bamboo forests.
				
				$N_{bamboo}(per\ plant)$: Number of bamboo plants.
				
				$(CF)_{bamboo}$: Carbon content of bamboo forest, taken as 0.5.
				
				\vspace{2mm}
				\noindent{$\divideontimes$ \bfseries Carbon Stocks of Shrublands}
				
				Carbon stocks of shrublands is the product of shrublands biomass and carbon content. The calculation formula is:
				\begin{equation}
					C_{shrublands} = W_{shrublands} \cdot A_{shrublands} \cdot (CF)_{shrublands}
				\end{equation}
				
				$C_{shrublands}$: Carbon stock of shrublands.
				
				$W_{shrublands}(t \cdot hm^{-2})$: Biomass per unit area of shrublands, average value is $19.76\ t \cdot hm^{-2}$.
				
				$A_{shrublands}(hm^2)$: Area of shrublands.
				
				$(CF)_{shrublands}$: Carbon content of shrublands, taken as 0.50.
			\subsubsection{Solutions to the Model}
				Similarly, taking Gansu as an example, in order to calculate the carbon stock of forest vegetation in the forest reserve, we collected data from some articles and aggregate carbon stock data which is showing in Table \ref{tab6}:
				\begin{table}[H]
					\centering
					\caption{Carbon Stock in Different Forests}
					\label{tab6}
					\begin{tabular}{p{2.2cm}<{\centering}p{2.2cm}<{\centering}p{2.5cm}<{\centering}p{2.5cm}<{\centering}p{2.5cm}<{\centering}}
						\Xhline{2pt}
						Forest Type & Area($/hm^2$) & Carbon Stock($/t \times 10^4$) & Carbon Stock Ratio(\%) & Converted to $CO_2$($t \times 10^4$) \\
						\Xhline{1pt}
						\rowcolor{blue!20}
						Arbor Forest & 97645.67 & 582.64 & 97.97 & 2136.54 \\
						Economic Forest & 2860.31 & 3.19 & 0.54 & 11.70 \\
						\rowcolor{blue!20}
						Bamboo Forest & 1297.45 & 7.49 & 1.26 & 27.47 \\
						shrublands & 1422.54 & 1.41 & 0.24 & 5.17 \\
						\rowcolor{blue!20}
						Total & 103225.97 & 594.73 & 100.00 & 2180.87 \\
						\Xhline{2pt}
					\end{tabular}
				\end{table}
				According to Table \ref{tab6}, the carbon stock of forest vegetation in the forest reserve achieved 5,826,400 tons, which converts to $CO_2$ is 21,365,400 tons.
	
				Among the four types of forests mentioned above, the total carbon stock of arbor forest, for example, is composed of carbon stocks of different forest stands and different age groups, which is shown in Table \ref{tab10}:
				\begin{table}[H]
					\centering
					\caption{Carbon stock composition of different forest stands in arbor forest}
					\label{tab10}
					\begin{tabular}{p{2.2cm}<{\centering}p{2.2cm}<{\centering}p{2.5cm}<{\centering}p{2.5cm}<{\centering}p{2.5cm}<{\centering}}
						\Xhline{2pt}
						Forest Stand Type & Area($/hm^2$) & Carbon Stock($/t \times 10^4$) & Carbon Stock Ratio(\%) & Converted to $CO_2$($t \times 10^4$) \\
						\Xhline{1pt}
						\rowcolor{blue!20}
						Coniferous forest & 83024.20 & 5827.81 & 90.59 & 1935.48 \\
						Broadleaf forest & 13071.68 & 54.04 & 9.28 & 198.16 \\
						\rowcolor{blue!20}
						Coniferous and Broadleaf mixed forest & 1549.79 & 0.79 & 0.14 & 2.90 \\
						Total & 97645.67 & 582.64 & 97.97 & 2136.54 \\
						\Xhline{2pt}
					\end{tabular}
				\end{table}
				\begin{table}[H]
					\centering
					\caption{Carbon stock composition of different forest stands in arbor forest}
					\begin{tabular}{p{2.2cm}<{\centering}p{2.2cm}<{\centering}p{2.5cm}<{\centering}p{2.5cm}<{\centering}p{2.5cm}<{\centering}}
						\Xhline{2pt}
						Age Groups & Area($/hm^2$) & Carbon Stock($/t \times 10^4$) & Carbon Stock Ratio(\%) & Converted to $CO_2$($t \times 10^4$) \\
						\Xhline{1pt}
						\rowcolor{blue!20}
						Young Forests & 29441.72 & 132.60 & 22.76 & 486.24 \\
						Middle-aged Forests & 39152.72 & 217.15 & 37.27 & 796.29 \\
						\rowcolor{blue!20}
						Mature Forests & 19385.81 & 147.58 & 25.33 & 541.18 \\
						Overripe Forests & 8408.00 & 74.81 & 12.84 & 274.33 \\
						\rowcolor{blue!20}
						Total & 97645.67 & 582.64 & 100.00 & 2136.54 \\
						\Xhline{2pt}
					\end{tabular}
				\end{table}
				Substituting the above data into the carbon stock calculation formula and weighting the carbon stocks of different kinds of forests to sum up, we get the results that the carbon stock of forest vegetation in Gansu Province in 2020 is 397345.004567 million tons, converting to RMB 238,407,274,000 at the market price of carbon emission rights trading.
	
				At the same time, we can also conclude that the carbon stock of tree forests accounts for the largest proportion of the total carbon stock, up to 97.97\%; followed by bamboo forests and economic forests, accounting for 1.26\% and 0.54\% respectively; the smallest proportion is shrub forests, accounting for only 0.24\%. 
				
				In terms of different stands of tree forests, the largest proportion is coniferous forests, accounting for 90.59\%; broad-leaved forests are the second largest, accounting for 9.28\% of the total carbon stock; the smallest proportion is mixed-leaved forests, accounting for only 0.14\%. In terms of age group structure, the largest proportion of carbon stock in different age groups of tree forests is in middle-aged forests, accounting for 37.27\%; followed by near-mature and young forests, accounting for 25.33\% and 22.76\%, respectively; the carbon stock in mature forests is smaller, accounting for 12.84\% of the carbon stock in different age groups; and finally, over-mature forests, accounting for only 1.80\%.
				
				With the rapid growth of young and middle-aged forests, their carbon sequestration effect increases greatly.
				Therefore, it is important to cultivate and invest in the process of ecological reserve construction, especially in young and middle-aged forests. This will help us to accomplish the task of "carbon peaking and carbon neutral".
	\section{Model Application 1:}
		\subsection{Choose Geographical locations}
			To solve the problem, firstly, we collected the global ecological protection of the environment and used PM2.5 as an indicator to help us select the right Asia-Pacific region for the study\cite{bib5}\cite{bib6}.
			\begin{figure}[H]
				\centering
				\includegraphics[width=\textwidth, height=8cm]{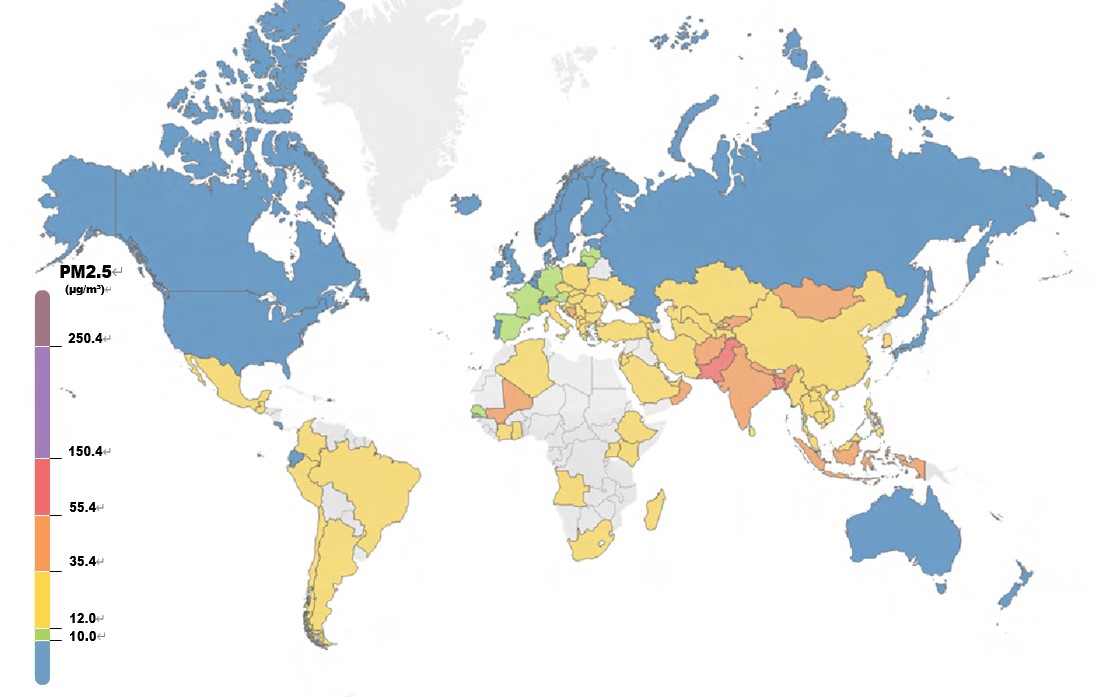}
				\caption{Global Map of Estimated PM2.5 Exposure by Countryside in 2020}
			\end{figure}
			From Fig \ref{fig4}, Asia, Africa and South America are the worst places for PM2.5 Exposure while Europe and North America are better than these continents.
			After analyzing the ranking data, in order to ensure the reasonableness of the simulation results, we choose Mongolia, which is ranked 3th in pollution and has a large area, as the subject of study.
		\subsection{Solutions to Model}
			\subsubsection{Description of the Ecological Environment of Mongolia}
				Based on our research object - Mongolia\cite{bib16}, we 
				collected data from 1975-2020 for the natural disaster-prone 
				Gobi Soumber province\cite{bib18}. Then, based on TASK 2, we 
				added $PM_{2.5}(pm_{2.5})$, $NO_x(NO)$, number of animals(NA), 
				number of microorganisms(NM) and number of plants(NP) to our 
				model. Some of these data are as follows:
				\begin{table}[H]
					\centering
					\caption{Raw data of Mongolia}
					\begin{tabular}{p{1cm}<{\centering}p{1cm}<{\centering}p{1cm}<{\centering}p{1cm}<{\centering}p{1cm}<{\centering}p{1cm}<{\centering}p{1cm}<{\centering}p{1cm}<{\centering}p{1cm}<{\centering}p{1cm}<{\centering}}
						\Xhline{2pt}
						DATE  & DEWP  & FRSHTT & GUST  & MXSPD & PRCP  & SLP   & SNDP  & VISIB & WDSP \\
						\Xhline{1pt}
						\rowcolor{blue!20}
						2020-6-1 & 33.3  & 0     & 999.9 & 17.5  & 0.00  & 1007.8 & 999.9 & 31.1  & 10.3 \\
						2020-6-2 & 41.5  & 0     & 999.9 & 11.7  & 0.00  & 1006.8 & 999.9 & 27.3  & 7.3 \\
						\rowcolor{blue!20}
						2020-6-3 & 36.4  & 0     & 999.9 & 11.7  & 0.01  & 1007.2 & 999.9 & 31.1  & 8.0 \\
						2020-6-4 & 41.0  & 10010 & 999.9 & 11.7  & 0.16  & 1005.4 & 999.9 & 17.6  & 7.1 \\
						\rowcolor{blue!20}
						2020-6-5 & 43.4  & 10000 & 999.9 & 11.7  & 0.03  & 997.9 & 999.9 & 26.1  & 5.3 \\
						2020-6-6 & 28.4  & 0     & 999.9 & 11.7  & 0.08  & 1010.6 & 999.9 & 31.1  & 8.5 \\
						\rowcolor{blue!20}
						2020-6-7 & 28.4  & 0     & 999.9 & 15.5  & 0.00  & 1016.4 & 999.9 & 28.0  & 9.0 \\
						2020-6-8 & 33.8  & 10000 & 999.9 & 7.8   & 0.28  & 1018.6 & 999.9 & 20.8  & 4.2 \\
						\rowcolor{blue!20}
						2020-6-9 & 37.0  & 0     & 999.9 & 5.8   & 0.00  & 1013.9 & 999.9 & 24.9  & 4.2 \\
						2020-6-10 & 29.8  & 10000 & 999.9 & 13.6  & 0.02  & 1014.0 & 999.9 & 27.5  & 7.0 \\
						\Xhline{2pt}
					\end{tabular}
				\end{table}
				Aftering reading the articles we collected\cite{bib14,bib15}, 
				we obtained the formula to calculate Ecological hazard 
				index(EH):
				\begin{equation}
					\begin{aligned}
						EH =& 0.246u + 0.2p + 0.04\Delta P_3 + 0.051t + 0.148\Delta t + 0.208dv + 0.019\Delta t_{24} + 0.072tr \\&+ 0.017u^3 + 0.003(pm_{2.5} + NO_x) + 0.12NA + 0.03NM + 0.14NP
					\end{aligned}
					\label{equ3}
				\end{equation}
				then, by calculation, we obtained the statistics for the period 1975-2021. Here, we list the data for the last 10 years as follows:
				\begin{table}[H]
					\centering
					\caption{Data calculation results for the last 10 years}
					\label{tab4}
					\begin{tabular}{p{1.5cm}<{\centering}p{2cm}<{\centering}p{2cm}<{\centering}p{2cm}<{\centering}p{2cm}<{\centering}p{2cm}<{\centering}}
						\Xhline{2pt}
						Year & 2012 & 2013 & 2014 & 2015 & 2016 \\
						\Xhline{1pt}
						\rowcolor{blue!20}
						Results & 34.3782 & 19.6250 & 27.8284 & 26.2017 & 24.4943 \\
						\Xhline{2pt}
					\end{tabular}
				\end{table}
				\vspace{-2.5em}
				\begin{table}[H]
					\centering
					\begin{tabular}{p{1.5cm}<{\centering}p{2cm}<{\centering}p{2cm}<{\centering}p{2cm}<{\centering}p{2cm}<{\centering}p{2cm}<{\centering}}
						\Xhline{2pt}
						Year & 2017 & 2018 & 2019 & 2020 & 2021 \\
						\Xhline{1pt}
						\rowcolor{blue!20}
						Results & 23.7349 & 20.8231 & 27.8250 & 22.3012 & 24.0422\\
						\Xhline{2pt}
					\end{tabular}
				\end{table}
				Accoring to Table \ref{tab4}, the ecological risk index for each of the last 10 years is higher than the ecological risk alert line 20. It illustrates that Mongolia's ecological environment is at a low level, so there is an urgent need to establish nature reserves.
				Finally, we chose the Central Gobi Province as the site for the establishment of the nature reserve.
			\subsubsection{Solutions}
				So, firstly, we collected data from 20 meteorological stations in Mongolia. Then, we used the AKQI meteorological station as an example to quantitatively evaluate the impact on the ecological environment before and after the establishment of the IKH Nature Reserves with Equation \ref{equ3}.
				
				Aftering collecting the data of IKH natural reserve(shown in Table \ref{tab9}), we calculate each indicator.
				\begin{table}[H]
					\centering
					\caption{Data of IKH Natural Reserve}
					\begin{tabular}{p{1.2cm}<{\centering}p{1.2cm}<{\centering}p{1cm}<{\centering}p{1cm}<{\centering}p{1cm}<{\centering}p{1cm}<{\centering}p{1.2cm}<{\centering}p{1.2cm}<{\centering}p{1.2cm}<{\centering}}
						\Xhline{2pt}
						\multicolumn{1}{c}{DATE} & DEWP & VISIB & WDSP  & denta\_T & tr    & pm2.5 & No\_x & score \\
						\Xhline{1pt}
						\rowcolor{blue!20}
						2020-6-20 & 46.6 & 31.1  & 12.9  & 25.7  & 5.7   & 139.268 & 0.058501 & 15.48997 \\
						2020-6-21 & 54.6 & 31.1  & 16    & 28.9  & -0.1  & 181.6823 & 0.111689 & 9.022184 \\
						\rowcolor{blue!20}
						2020-6-22 & 50.8 & 31.1  & 8     & 32.6  & 11.5  & 155.3766 & 0.109386 & 11.60402 \\
						2020-6-23 & 44.4 & 31.1  & 11.4  & 30.2  & 1.8   & 100.1941 & 0.078607 & 12.1231 \\
						\rowcolor{blue!20}
						2020-6-24 & 45.3 & 26.9  & 12.4  & 29.2  & 3.8   & 190.0031 & 0.088087 & 13.60865 \\
						2020-6-25 & 49.3 & 31.1  & 11.7  & 29.5  & 1.8   & 132.0534 & 0.089902 & 8.311942 \\
						\rowcolor{blue!20}
						2020-6-26 & 46.3 & 31.1  & 6.1   & 34.6  & 1.8   & 134.2819 & 0.130479 & 19.51446 \\
						2020-6-27 & 42 & 31.1  & 8.3   & 32.8  & 3.8   & 75.76843 & 0.102386 & 9.440927 \\
						\rowcolor{blue!20}
						2020-6-28 & 44.2 & 31.1  & 10.4  & 29.2  & 3.8   & 118.9839 & 0.064043 & 14.09339 \\
						2020-6-29 & 51.2 & 28    & 8.7   & 30.4  & 3.8   & 118.7926 & 0.071861 & 8.295134 \\
						\Xhline{2pt}
					\end{tabular}
					\label{tab9}
				\end{table}
				In this case, for the visual impairment index, we use the formula:
				\begin{equation}
					DV = dv + x_1
				\end{equation}
				The Visibility Indicators were calculated for each year and the weighted average was taken to obtain the following results:
				\begin{table}[H]
					\centering
					\caption{Visibility Indicators}
					\label{tab7}
					\begin{tabular}{p{1.5cm}<{\centering}p{2cm}<{\centering}p{2cm}<{\centering}p{2cm}<{\centering}p{2cm}<{\centering}p{2cm}<{\centering}}
						\Xhline{2pt}
						Year & 2017 & 2018 & 2019 & 2020 & 2021 \\
						\Xhline{1pt}
						\rowcolor{blue!20}
						Results & 10.3585 & 9.4106 & 8.5184 & 9.3673 & 9.2684\\
						\Xhline{2pt}
					\end{tabular}
				\end{table}
				\begin{figure}[H]
					\subfigure[Visibility Index]{
						\begin{minipage}[H]{0.4\textwidth}
							\centering
							\includegraphics[width=7cm, height=4.2cm]{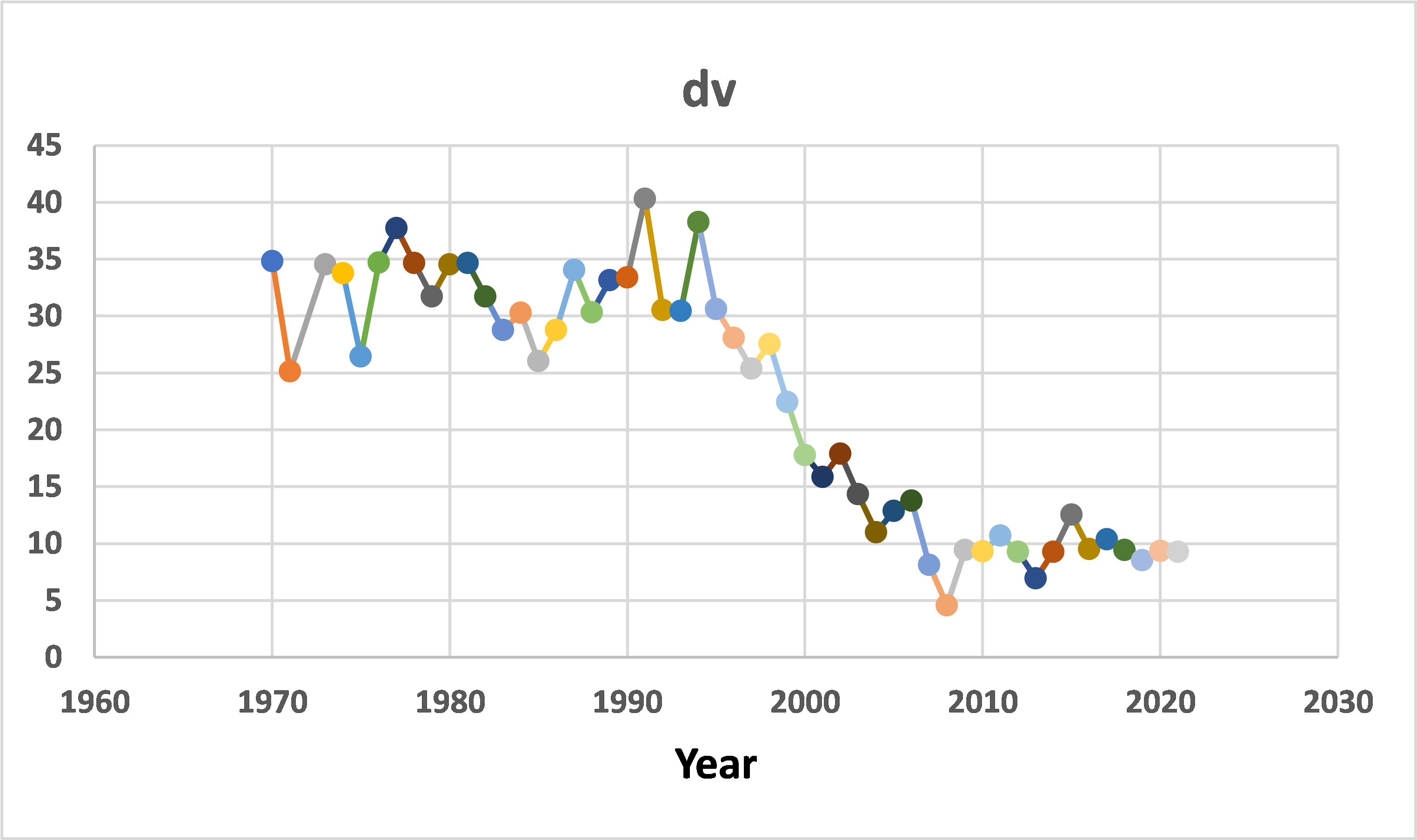}
							\label{fig4}
					\end{minipage}}
					\hspace{10mm}
					\subfigure[Sandstorm Development Trend Indicator]{
						\begin{minipage}[H]{0.4\textwidth}
							\centering
							\includegraphics[width=7cm, height=4.2cm]{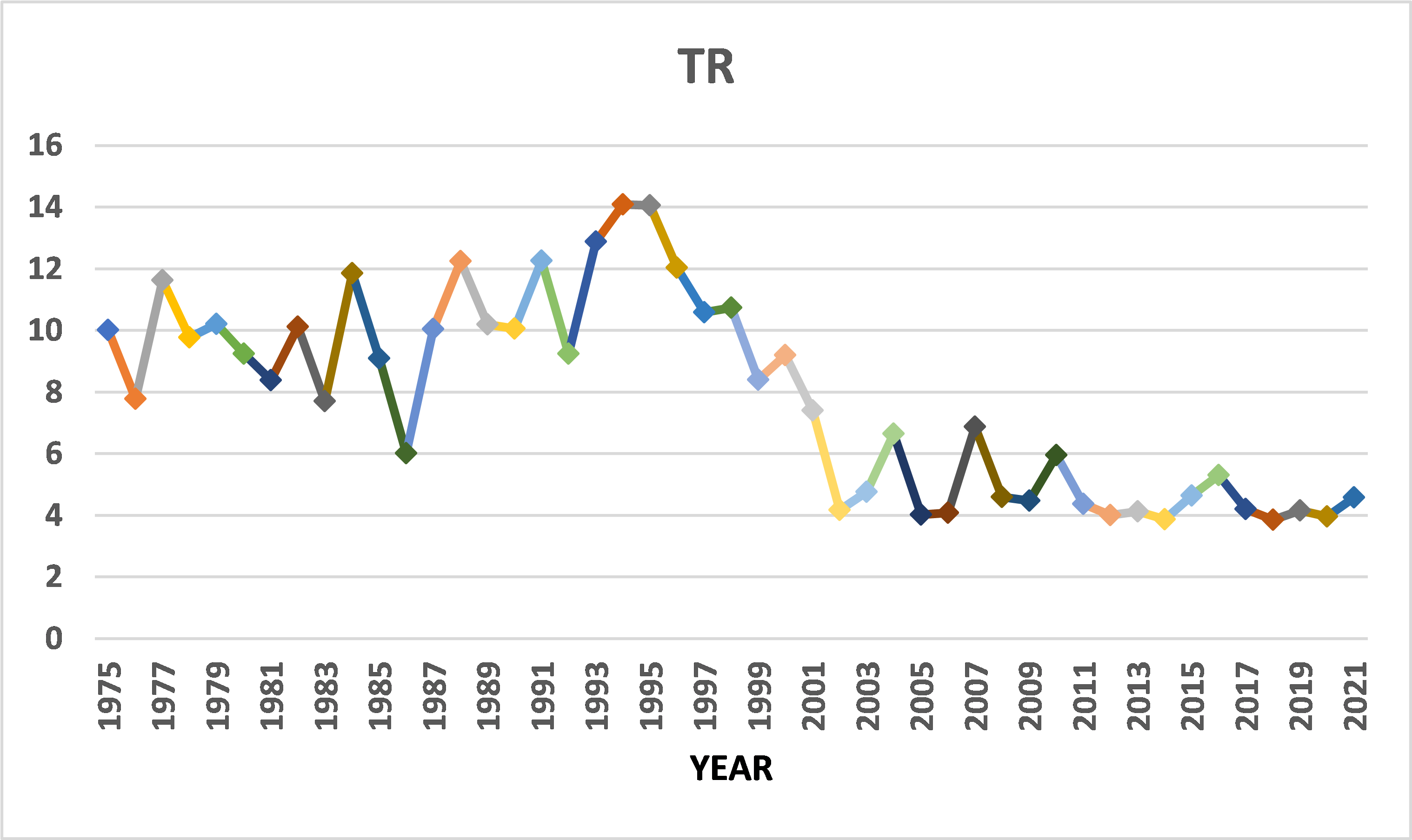}
							\label{fig5}
					\end{minipage}}
					\caption{Indicators}
				\end{figure}
				As we can see from Table \ref{tab7} and Fig \ref{fig4}, in recent years, the visibility index of Mongolia showed a decreasing trend at the beginning of the 21st century and remained relatively stable after 2010, which to some extent can indicate that the ecological environment of Mongolia is changing for the better.
	
				For the Sandstorm Development Trend Indicator , we use the formula:
				\begin{equation}
					\begin{aligned}
						TR &= TR + X_4 \\
						&= u - u_s + x_4
					\end{aligned}
				\end{equation}
				
				The Sandstorm Development Trend Indicator were calculated for each year and the weighted average was taken to obtain the following results:
				\begin{table}[H]
					\centering
					\caption{Sandstorm Development Trend Indicator}
					\label{tab8}
					\begin{tabular}{p{1.5cm}<{\centering}p{2cm}<{\centering}p{2cm}<{\centering}p{2cm}<{\centering}p{2cm}<{\centering}p{2cm}<{\centering}}
						\Xhline{2pt}
						Year & 2017 & 2018 & 2019 & 2020 & 2021 \\
						\Xhline{1pt}
						\rowcolor{blue!20}
						Results & 4.2036 & 3.8603 & 4.1532 & 3.9609 & 4.5725\\
						\Xhline{2pt}
					\end{tabular}
				\end{table}
				Likewise, 
				Using the similar approach, we end up with the values of 12 indicators for the Central Gobi Province for the period 1975-2021.
				Finally, we obtain the value of the risk index for 1975-2021(here we list 5 of them below):
				\begin{table}[H]
					\begin{tabular}{p{1.5cm}<{\centering}p{2cm}<{\centering}p{2cm}<{\centering}p{2cm}<{\centering}p{2cm}<{\centering}p{2cm}<{\centering}}
						\Xhline{2pt}
						Year & 2017 & 2018 & 2019 & 2020 & 2021 \\
						\Xhline{1pt}
						\rowcolor{blue!20}
						Risk & 10.9583 & 9.7789 & 10.5084 & 10.6718 & 12.3576\\
						\Xhline{2pt}
					\end{tabular}
				\end{table}
				\begin{figure}[H]
					\centering
					\includegraphics[width=\textwidth, height=8cm]{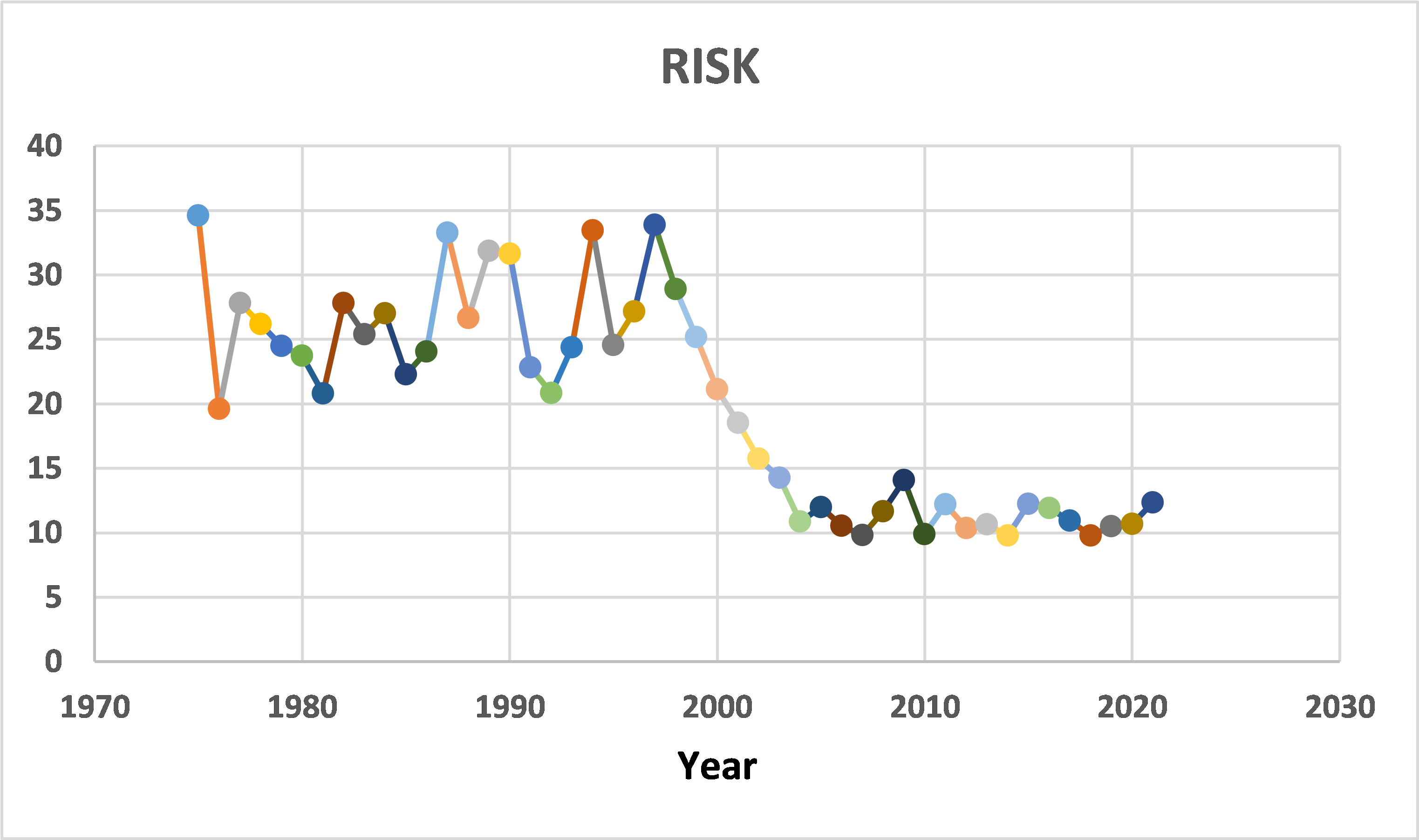}
					\caption{The Risk Index}
				\end{figure}
				It can be seen that before the establishment of the IKH Nature Reserve in 1996, the risk index of the Central Gobi Province was 27.15, which was higher than the ecological risk threshold of 20. After the establishment of the IKH Nature Reserve, the risk index of the Central Gobi Province was reduced to 12.357 after 14 years of protection. The area of the Central Gobi province is 178,000 km² and the area of the ikh nature reserve is 66,600 hectares. Combined with the formula, we can calculate $EH$ as:
				\begin{equation}
					\Delta EH = 178,000 \times (27.15-12.357)/66,600 = 39.537
				\end{equation}
				which reduces the risk index of the region by 39.537 per unit area of the nature reserve for 10 years.
				
				We counted the occurrence of natural disasters in the Central Gobi Province, to which the IKH Nature Reserve belongs, from 1975 to 2021 as follows:
				\begin{figure}[H]
					\centering
					\includegraphics[width=\textwidth, height=7cm]{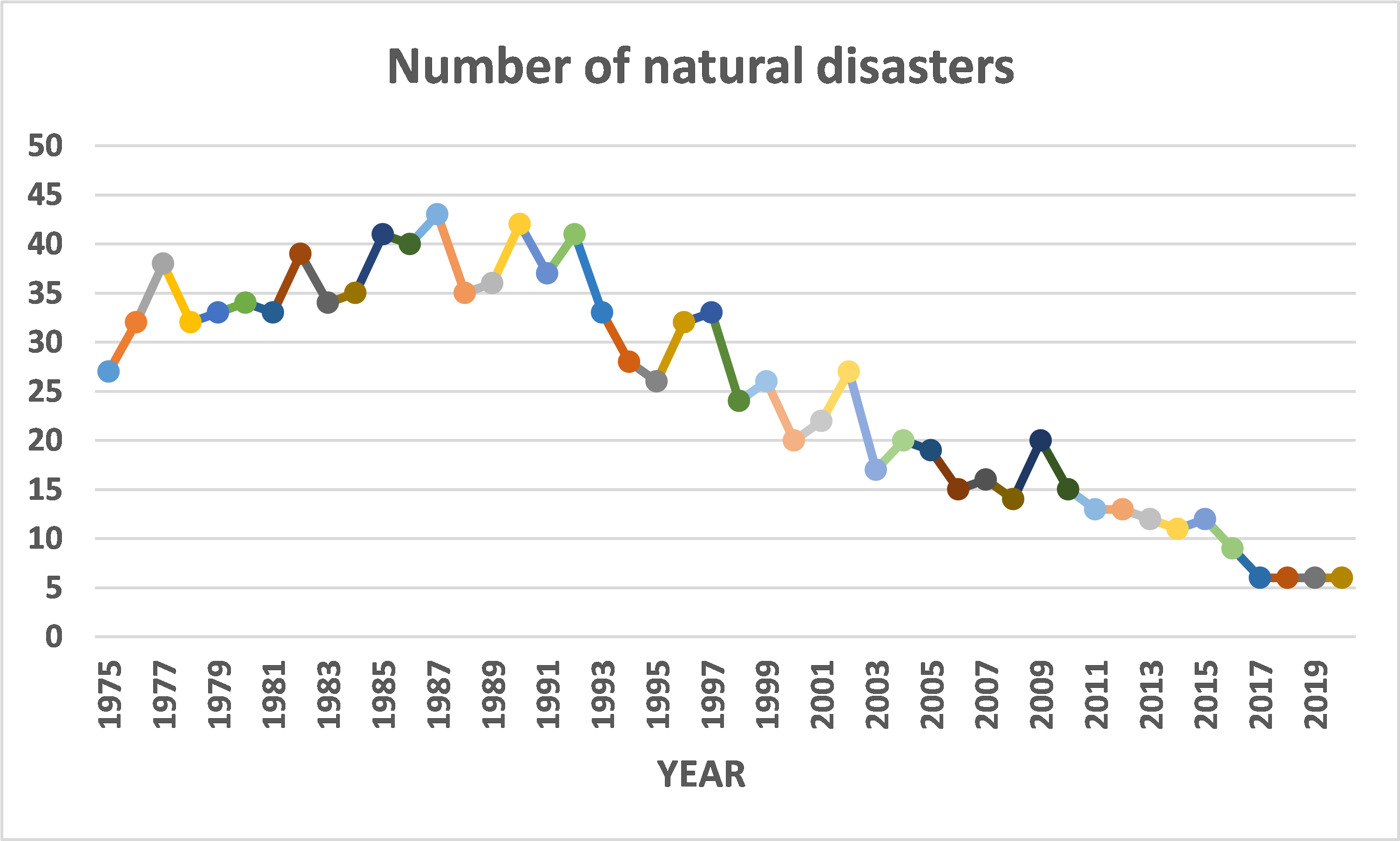}
					\caption{The Risk Index}
				\end{figure}
				It can be seen that after the establishment of the IKH nature reserve in 1996, the number of natural disasters in the Central Gobi Province has steadily decreased due to the better ecological environment and increased resistance to natural disasters, which directly reflects the benefits of the establishment of the nature reserve in building the ecological environment.
		\subsection{Evaluate Its Impact on Absorbing Greenhouse Gases and Mitigating Carbon Emissions}
			Following the calculation of carbon uptake by forest reserves in TASK 3, we have the following analytical data:
			
			At the same time, we can also conclude that the carbon stock of tree forests accounts for the largest proportion of the total carbon stock, up to 95.36\%; followed by bamboo forests and economic forests, accounting for 2.14\% and 1.55\% respectively; the smallest proportion is shrub forests, accounting for only 0.95\%. 
			
			In terms of different stands of tree forests, the largest proportion is coniferous forests, accounting for 91.23\%; broad-leaved forests are the second largest, accounting for 5.21\% of the total carbon stock; the smallest proportion is mixed-leaved forests, accounting for only 3.56\%. In terms of age group structure, the largest proportion of carbon stock in different age groups of tree forests is in middle-aged forests, accounting for 41.19\%; followed by near-mature and young forests, accounting for 24.62\% and 21.37\%, respectively; the carbon stock in mature forests is smaller, accounting for 10.24\% of the carbon stock in different age groups; and finally, over-mature forests, accounting for only 2.58\%.
			
			Substituting the above data into the carbon stock calculation formula and weighting the carbon stocks of different kinds of forests to sum up, we get the results that the carbon stock of forest vegetation in Gansu Province in 2020 is 160778.776 million tons, converting to RMB 964,672,669,900 at the market price of carbon emission rights trading.
	\section{Model Application 2:}
		\subsection{Current Status}
			Three indicators were used to model the impact of the model on the regional ecological environment: the ecological condition index, the dust storm hazard index and the forest carbon stock\cite{bib7}. In general, the ecological condition index and the dust storm hazard index are used to assess the need for a particular region to establish the Saihanba ecological model, while the forest carbon stock is used to assess the impact of establishing the ecological model on the carbon neutrality target.
			
			In China, the area with a low ecological condition index and a high dust storm hazard index is Gansu Province, which we found and obtained data from {\bfseries{balabala}}. The data were obtained from {\bfseries balabala}  and the ecological condition index was calculated as {\bfseries balabala}. and the dust storm hazard index is {\bfseries{balabala}}. We determined that it needs to establish the ecological model of Saihanba, and then calculated its forest carbon stock after establishing the ecological model as 397345.004567 million tons, which is equivalent to RMB 23840700.274 million at the market price of carbon emission rights trading, and has the most significant beneficial impact on the carbon neutral target.
		\subsection{Suggestions to Saihanba Ecological Model}
			A few suggestions for establishing the Sekhangba ecological model:
			
			(1) The forest area is expanded as much as possible, and the proportion of forest is mainly tree and coniferous forest, and the proportion of near-mature forest is increased by reasonable tree planting.
			
			(2) Control carbon emissions from forest edge areas.
			
			(3) Introduce as many species as possible within reasonable limits to increase biodiversity.
	
			(4) Reduce industrial waste water and gas emissions.
			
			(5) The ecological model should be established in dust-prone areas.
			
			If the ecological model construction is developed naturally according to the existing condition, although the comprehensive ecological level is improved, the coordination of the system still has problems, especially the impact on resources and environment is intensified.So we should:
			
			\vspace{2mm}
			\noindent{$\divideontimes$ \bfseries Adjust the industrial structure according to the location advantage.}
			
			Restrict the use of coal, especially the use of coal in the electric power industry, not in the new coal-fired power plants, to maximize the potential of renewable energy around, and replace coal-fired power generation with renewable energy generation.
			
			Shut down, rectify cement plants, steel plants and other high-pollution, high-energy industries, and vigorously promote gas boilers instead of coal-fired
			coal-fired boilers.
			
			Comprehensive transformation of civilian and small-scale commercial stoves, so that civilian fuel from coal to gas conversion, the ban on agricultural waste burning.
	
			\vspace{2mm}
			\noindent{$\divideontimes$ \bfseries Moderate urban development, focus on urban quality.}
			
			Choosing a moderate rate of urban development.
			
			Develop a scientific scale of urban development
			
			\vspace{2mm}
			\noindent{$\divideontimes$ \bfseries Establish an effective mechanism to implement intensive savings.}
			
			It should unify emission standards and environmental quality standards as soon as possible, clarify the overall objectives and index system of environmental protection, carry out the delineation of regional ecological protection red line and environmental quality red line according to the main function to delineate, promote regional emissions and other aspects of unified environmental protection and construction planning, unified emission standards, unified oil use standards, unified governance standards, and at the same time carry out the development of integrated environmental protection planning. Repeatedly advocate and guide low-carbon life by widely using media means.
			
			\vspace{2mm}
			\noindent{$\divideontimes$ \bfseries Control environmental pollution. Pollution, improve environmental quality.}
	
			Take the circular economy as a way to achieve the recycling of pollutants. Vigorously develop the circular economy, change the past "mass production
			Production, consumption, waste" the traditional growth model, to "reduce, reuse, resource" as the principle, to low consumption, low emissions, high efficiency as the basic features, to achieve waste reduction, resource and harmless.
			With low consumption, low emission, high efficiency as the basic characteristics, to achieve waste reduction, resourcefulness and harmlessness. Following the example of the United States, Germany
			The United States, Germany and other developed countries, clear producer responsibility, the development of economic incentives; promote manufacturers to recycle product packaging, or even at the end of the product life of the recycling products; vigorously promote the recycling of products.
			The product is recycled at the end of its useful life; the comprehensive use of waste generation, recycling
			We should vigorously promote the comprehensive utilization in the waste generation process and recycle various waste resources generated in the social consumption process; we should vigorously promote green consumption to make the economic
			system and the natural ecological system, and maintain the natural ecological balance, which is based on the efficient use and recycling of resources.
			The core of the utilization is the efficient use and recycling of resources.
			
			\vspace{2mm}
			\noindent{$\divideontimes$ \bfseries Protect the ecological environment and develop ecological culture.}
			
			According to different ecological function areas, carry out ecological protection and construction. According to the ecological environment characteristics, ecological
			Sensitivity and ecological service functions of different types of functional areas, and carry out different protection measures. Improve the green space system
			To improve the green space system and create a suitable living environment. Extend the concept of environmental protection and highlight the ecological culture. The balance of natural ecological environment is the most basic guarantee for human
			The balance of natural ecological environment is the most basic guarantee for sustainable development, while the history, culture and national culture are the soul and vitality of human development.
	\section{Analysis on Model's Sensitivity}
		\subsection{Sensitivity Analysis of the Risk Degree of Sandstorm(H)}
			In our model, we have created two indicators, the ecological condition index and the dust storm risk. Next, we perform a sensitivity analysis of the Rish Degree of Sandstorm(H).
			
			As we demonstrated in TASK 2, the composite indicator of Risk Degree of Sandstorm(H) can be calculated as:
			\begin{equation}
				\begin{aligned}
					H &= f(DV, U, \Delta T, TR, P) \\
					&= 0.159 DV + 0.403 U + 0.088 \Delta T + 0.077 TR + 0.274 P
				\end{aligned}
			\end{equation}
			
			Suppose that for a particular object, we calculate its dust storm hazard index:
			\begin{equation}
				H = 0.246u + 0.2p + 0.04p_3 + 0.051t + 0.148\Delta t + 0.208dv + 0.019\Delta t_{24} + 0.072tr + 0.017u^3
			\end{equation}
			Take the principal components u and p for sensitivity analysis:
			\begin{equation}
				\frac{\Delta H}{\Delta u} = 0.246 + 0.051u^2
			\end{equation}
			when $\Delta u$ equals to 0.1$u$:
			\begin{equation}
				\Delta H = 0.0246u + 0.0051u^3 \approx 0.0246u
			\end{equation}
			Thus the model is robust to u.
	
			Going on, we assumed that $\Delta p_3 \approx k\Delta p$, then
			\begin{equation}
				\Delta H = 0.2 \Delta p + 0.04 \Delta p^2 \approx 0.2\Delta p
			\end{equation}
			Thus the model is robust to p.
			
			Through the sensitivity test we found that the effect of parameter changes on the model results is within our acceptable range, and therefore the model we have built is stable.

\end{document}